\documentclass[fleqn]{mnras}

\usepackage{newtxtext}
\usepackage{newtxmath}
\usepackage{esint}

\usepackage{bigints}

\usepackage{float}
\usepackage{tikz}
\usepackage{graphicx}
\usepackage{times}
\usepackage{mathtools}
\usepackage{amsmath}
\usepackage{verbatim}
\usepackage{bm}

\usepackage{overpic}

\usepackage{yfonts}

\pdfoutput=1

\catcode`\@=11
\newcommand\gsim{\ifmmode{\mathrel{\mathpalette\@versim>}}
    \else{$\mathrel{\mathpalette\@versim>}$}\fi}
\newcommand\lsim{\ifmmode{\mathrel{\mathpalette\@versim<}}
    \else{$\mathrel{\mathpalette\@versim<}$}\fi}
\catcode`\@=12

\DeclareMathAlphabet{\mathpzc}{OT1}{pzc}{m}{it}

\DeclareMathAlphabet{\mathcal}{OMS}{cmsy}{m}{n}% per il \MR

\newcommand\xv{{\bf x}}

\newcommand{\tx}{\tilde{x}}
\newcommand{\ty}{\tilde{y}}
\newcommand{\tz}{\tilde{z}}

\newcommand{\arccot}{{\rm arccot}}
\newcommand{\arcsh}{{\rm arcsinh}\hspace{0.5mm}}

\newcommand\rhotil{\tilde\rho}
\newcommand\rtz{\rho_0}
\newcommand\rtu{\rho_1}
\newcommand\rtd{\rho_2}

\newcommand\vrtz{\varrho_0}
\newcommand\vrtu{\varrho_1}
\newcommand\vrtd{\varrho_2}

\newcommand\rhon{\rho_{\rm n}}

\newcommand{\vcirc}{\varv_{\rm c}}

\newcommand\Rtil{\tilde{R}}
\newcommand\ztil{\tilde{z}}

\newcommand\PsiT{\Psi_{\rm T}}

\newcommand\Psin{\Psi_{\rm n}}

\newcommand\tPsiz{\Psi_0}
\newcommand\tPsiu{\Psi_1}
\newcommand\tPsid{\Psi_2}

\newcommand\tpsiz{\psi_0}
\newcommand\tpsiu{\psi_1}
\newcommand\tpsid{\psi_2}
\newcommand\tpsii{\psi_i}

\newcommand\Ms{M_*}
\newcommand\Mbh{M_{\rm {BH}}}

\newcommand\sigBH{\sigma_{\rm BH}}
\newcommand\sigs{\sigma_*}
\newcommand\sigphi{\sigma_{\varphi}}
\newcommand\vphib{\overline{\varv_\varphi}}

\newcommand\vlos{\varv_{\rm los}}
\newcommand\siglos{\sigma_{\rm los}}

\newcommand\Dels{\Delta_*}
\newcommand\DelBH{\Delta_{\rm BH}}

\newcommand\vphi{\varv_{\varphi}}
\newcommand\vz{\varv_z}

\newcommand\vR{\varv_R}

\newcommand\tvcz{\varv^2_0}
\newcommand\tvcu{\varv^2_1}

\newcommand{\ellE}{\mathbb{E}}
\newcommand{\ellF}{\mathbb{F}}

\newcommand{\Hij}{H_{ij}}
\newcommand{\Hzz}{H_{00}}
\newcommand{\Hzu}{H_{01}}
\newcommand{\Hzd}{H_{02}}

\newcommand{\Huu}{H_{11}}
\newcommand{\Hud}{H_{12}}

\newcommand{\Hdd}{H_{22}}

\newcommand{\Xij}{X_{ij}}

\newcommand{\Zij}{Z_{ij}}
\newcommand{\Zzd}{Z_{02}}
\newcommand{\Zud}{Z_{12}}
\newcommand{\Zdd}{Z_{22}}

\title[Jeans modelling of weakly flattened stellar systems]{Jeans modelling of weakly flattened ellipsoidal systems}

\author[A. Mancino, L. Ciotti, S. Pellegrini \& F. Giannetti]{Antonio Mancino$^{1,2}$, Luca Ciotti$^1$, Silvia Pellegrini$^{1,2}$ \& Federica Giannetti$^1$
\\
$^1$Department of Physics and Astronomy, University of Bologna, via Gobetti 93/3, 40129 Bologna, Italy
\\
$^2$Istituto Nazionale di Astrofisica (INAF), Osservatorio di Astrofisica e Scienza dello Spazio di Bologna (OAS), Via Gobetti 93/3, Bologna 40129, Italy}

\date{Accepted 2023 December 9. Received 2023 December 8; in original form 2023 November 16}

\begin{document}
\maketitle
%\label{firstpage}

  %%%%%%%%%%%%%%%%%%%%%%%%%%%%%%%%%%%%%%%%%%%%%%%%%%%%%%%%%%%%%%%%%%%%%%%%%%%%%%%%%%%%%%%%%%%%%%
  %*******************************************SOMMARIO******************************************
  %%%%%%%%%%%%%%%%%%%%%%%%%%%%%%%%%%%%%%%%%%%%%%%%%%%%%%%%%%%%%%%%%%%%%%%%%%%%%%%%%%%%%%%%%%%%%%

  \begin{abstract}
  In the homoeoidal expansion, a given ellipsoidally stratified density distribution,
  and its associated potential, are expanded in the (small) density flattening parameter 
  $\eta$, and usually truncated at the linear order. The truncated density-potential pair
  obeys exactly the Poisson equation, and it can be interpreted as the first-order 
  expansion of the original ellipsoidal density-potential pair, or as a new autonomous
  system. In the first interpretation, in the solutions of the Jeans equations the 
  quadratic terms in $\eta$ must be discarded (``$\eta$-linear'' solutions), while in the
  second (``$\eta$-quadratic'') all terms are retained. In this work 
  we study the importance of the quadratic terms by using the ellipsoidal Plummer model 
  and the Perfect Ellipsoid, which allow for fully analytical $\eta$-quadratic solutions. 
  These solutions are then compared with those obtained numerically for the original 
  ellipsoidal models, finding that the  $\eta$-linear models already provide an excellent 
  approximation of the numerical solutions. As an application, the $\eta$-linear Plummer 
  model (with a central black hole) is used for the phenomenological interpretation of the 
  dynamics of the weakly flattened and rotating globular cluster NGC 4372,
  confirming that this system cannot be interpreted as an isotropic rotator,  
  a conclusion reached previously with more sophisticated studies. 
  \end{abstract}


  %%%%%%%%%%%%%%%%%%%%%%%%%%%%%%%%%%%%%%%%%%%%%%%%%%%%%%%%%%%%%%%%%%%
  \section{Introduction}\label{sec:Intro}
  %%%%%%%%%%%%%%%%%%%%%%%%%%%%%%%%%%%%%%%%%%%%%%%%%%%%%%%%%%%%%%%%%%%
  
  For their intrinsic simplicity in stellar dynamics, spherical models are of great 
  utility both in theory and applications (e.g. Binney \& Tremaine 2008, hereafter BT08; 
  Bertin 2014; Ciotti 2021, hereafter C21).  However, phenomena such as rotation can be 
  studied properly only by allowing for some flattening in the density profile.
  Among the specific difficulties encountered in modelling non-spherical systems, the 
  derivation of the gravitational potential is certainly a major one, and, apart from 
  special cases in which the solution is known in explicit form, one must resort to 
  time-consuming numerical integrations. In turn, if the potential is only known 
  numerically, then also the Jeans equations must be solved numerically, making the 
  exploration of the parameter space even more laborious. Fortunately, many real stellar 
  systems are characterised by small deviations from spherical symmetry; examples of 
  weakly flattened systems are, for instance, E1/E3 galaxies, and many Globular Clusters  
  (e.g. Varri \& Bertin 2012). For these systems, the technique of the homoeoidal expansion 
  (Ciotti \& Bertin 2005, hereafter CB05) provides an approximate, yet robust, and easy 
  procedure to build one- and multi-component dynamical models (Ciotti et al. 2021, 
  hereafter CMPZ21; see also Chapter 13 in C21, and references therein). 
  
  In practice, in the homoeoidal expansion method, a chosen ellipsoidally stratified 
  density distribution, and the associated potential, are usually expanded and truncated 
  at the linear order in terms of the density flattening $\eta$, therefore producing, 
  thanks to the linearity of Poisson's equation, an exact density-potential pair. 
  Both the density and the potential are written as a spherical part 
  plus a non-spherical term proportional to $\eta$; the non-spherical term, in turn, 
  reduces to the product of the square of the cylindrical radius $R$ and a spherical 
  function. This very specific structure allows for manageable solutions of the Jeans 
  equations. Of course, the truncation at the linear order in $\eta$ of the original 
  density-potential pair is only matter of convenience, as the linearity of the Poisson 
  equation implies that, at any truncation order in $\eta$, the resulting truncated 
  functions are an exact density-potential pair. However, increasing the order of 
  truncation increases also the number of terms to be considered in the solution of the 
  Jeans equations; therefore it is  natural to ask how good the linear truncation already 
  is, in order to save as much computational effort as possible,  while maintaining a 
  reasonable description of the original system.
  
  A second closely related question arises when considering the solution of the Jeans 
  equations even for the linearly truncated density-potential pair. 
  In fact, the truncated density-potential pair can be seen in two different ways: as 
  the first-order expansion of the ellipsoidal 
  parent galaxy model in the limit of small flattening, or as an independent non-spherical system.  
  In the first interpretation (``$\eta$-linear''), only linear terms in the flattening 
  are retained in the solution of the Jeans equations; in the second interpretation
  (``$\eta$-quadratic''), the Jeans equations contain up to quadratic terms in the flattening. 
  In previous works, only the 
  first interpretation has been discussed; here we investigate the effects of including 
  higher order $\eta$-terms in the solutions of the Jeans equations: it is quite obvious 
  that the discarded terms, proportional to $\eta^2$ and depending on coordinates, are not 
  necessarily small, and in principle in some regions of space they could be even larger 
  than lower order terms\footnote{
  As an example, consider the quadratic truncation $1-\eta x+\eta^2 x^2/2$ of the uniformly 
  convergent expansion of ${\rm e}^{-\eta x}$.}.
  
  In this paper we address these questions, and compare the solutions of the Jeans 
  equations of the original ellipsoidal model (hereafter ``full solution'') with their 
  $\eta$-linear and $\eta$-quadratic expansions, in order to quantify the performance of 
  the homoeoidal expansion method. For this study, we consider two simple ellipsoidal 
  models: the Perfect Ellipsoid (de Zeeuw \& Lynden-Bell 1985, hereafter ZL85) and the 
  ellipsoidal Plummer (1911, hereafter P11) model, to which a central
  black hole (BH) is added. For these, the $\eta$-quadratic (and 
  so the $\eta$-linear) solutions can be evaluated analytically in terms of elementary 
  functions. In particular, for each model we compare the two expanded solutions with that
  recovered numerically for the original model. We find that the $\eta$-linear approximation
  suffices to provide excellent agreement with the numerical solution, for small but realistic 
  values of $\eta$. We then exploit this result by building an $\eta$-linear Plummer model 
  for the Globular Cluster (GC) NGC 4372, to investigate the 
  relation between flattening and ordered rotation. This GC is a natural candidate for an 
  exploratory study with our method, since its density profile is well described by the Plummer
  model (see Kacharov et al. 2014), and its flattening is sufficiently low for applicability 
  of the  $\eta$-linear modeling. Another motivation for this application is given by the 
  remarkable property of homoeoidally expanded systems of having a streaming velocity field 
  that scales as $\sqrt{\eta}$, increases linearly with radius in the central regions,
  and decreases outward after a maximum; these features are noticeably similar to those
  commonly observed, or adopted in a phenomenological description, for the rotation curves of 
  GCs. In agreement with previous results obtained with more sophisticated methods (e.g. Varri 
  \& Bertin 2012; Bianchini et al. 2013; Jeffreson et al. 2017), our simple modelling excludes
  the possibility that NGC 4372 is an isotropic rotator, and we suggest instead that the observed
  rotational structure is possibly due to the presence of a rotating stellar substructure.
  
  The paper is organised as follows.  In Section \ref{sec:HomExp} we summarize the technical 
  details and main formulae of the homoeoidal expansion. In Section \ref{sec:JEs}, we set up 
  and discuss the $\eta$-linear and $\eta$-quadratic solutions of the associated Jeans equations.
  In Section \ref{sec:Models}, the main structural and dynamical properties of the two families
  of models are presented, with a detailed analysis of the solutions of the Jeans equations 
  generated by the two different interpretations. Finally, in Section \ref{sec:NGC4372} we 
  present a simple application of the $\eta$-linear modelling to the GC NGC 4372.

  %%%%%%%%%%%%%%%%%%%%%%%%%%%%%%%%%%%%%%%%%%%%%%%%%%%%%%%%%%%%%%%%%%%%%%%
  \section{The Homoeoidal Expansion}\label{sec:HomExp}
  %%%%%%%%%%%%%%%%%%%%%%%%%%%%%%%%%%%%%%%%%%%%%%%%%%%%%%%%%%%%%%%%%%%%%%%
  
  We recall the main properties of the homoeoidal expansion. Let $\xv=(x,y,z)$, and 
  consider a mass density distribution $\rho(\xv)$ of total mass $M$ stratified on 
  ellipsoidal surfaces 
  \begin{equation}
         m^2 \equiv \frac{x^2}{a^2}+\frac{y^2}{b^2}+\frac{z^2}{c^2}
         =\tx^2+\frac{\ty^2}{q_y^2}+\frac{\tz^2}{q_z^2}\hspace{0.2mm},
  \label{eq:m}
  \end{equation} 
  with $a \geq b \geq c > 0$, $\tx \equiv x/a$, $\ty \equiv y/a$, $\tz \equiv z/a$, 
  $q_y \equiv b/a$, and $q_z \equiv c/a$. For $q_y=q_z=1$, the distribution is 
  spherically symmetric, the oblate case corresponds to $q_y=1$ and $0<q_z<1$, and 
  the prolate to $0<q_y=q_z<1$. We write  
  \begin{equation}
        \frac{\rho(\xv)}{\rhon}=\frac{\rhotil(m)}{q_y q_z}\hspace{0.2mm},
        \qquad 
        \rhon\equiv\frac{M}{4\upi a^3}.
  \label{eq:rho(x)}
  \end{equation}
  Notice that $\rhotil$, as a function of its argument, is independent of $q_y$
  and $q_z$, and it can be identified with the spherical member of the ellipsoidal
  family. Notice that the presence of the coefficient $q_y q_z$ at the denominator in 
  equation \eqref{eq:rho(x)} guarantees that $M$ is the mass of the model independently 
  of the flattening\footnote{For the normalization in case of an infinite total mass
  see C21.}, and, from equation \eqref{eq:rho(x)}, the mass within the ellipsoid $m$ 
  reads
  \begin{equation}
     \frac{M(m)}{M}=\int_0^m\rhotil(t)\hspace{0.4mm}t^2dt.
  \end{equation}
  The (relative) potential associated with the density distribution in equation 
  \eqref{eq:rho(x)} is given by 
  \begin{equation}
        \frac{\Psi(\xv)}{\Psin}=
        \frac{1}{4} \bigintsss_{\hspace{0.4mm}0}^\infty\hspace{-0.1mm}
        \frac{F[m(\xv\hspace{0.2mm},u)]}{\sqrt{\hspace{0.2mm} D(u)\hspace{0.3mm}}}\hspace{0.3mm}
        du,
        \qquad\,\,
        F(m) \equiv 2\hspace{-0.25mm}\int_m^\infty \rhotil(t)\hspace{0.3mm}t\hspace{0.1mm}dt,
  \label{eq:Chandra}
  \end{equation}
  (e.g. BT08; see also exercise 2.12 in C21), where $\Psin \equiv GM/a$, $G$ is the 
  gravitational constant,
  \begin{equation}
        D(u) \equiv 
        (1+u)\hspace{0.2mm}(q_y^2\hspace{-0.25mm}+u)\hspace{0.2mm}(q_z^2\hspace{-0.25mm}+u)
        \hspace{0.3mm},
  \end{equation}
  and
  \begin{equation}
         m^2(\xv\hspace{0.3mm},u) \equiv 
         \frac{\tx^2}{1+u}+\frac{\ty^2}{q_y^2+u}+\frac{\tz^2}{q_z^2+u}
         \hspace{0.3mm}.
  \label{eq:mu}
  \end{equation}
  Finally (Roberts 1962), the gravitational self-energy $W$ can be written as
  \begin{equation}
        \frac{W}{M\Psin}=
        -\hspace{0.4mm}\frac{\varw_1+q_y^2\hspace{0.4mm}\varw_2+q_z^2\hspace{0.4mm}\varw_3}{16\hspace{0.4mm}q_y q_z}        
        \hspace{0.25mm}
        \hspace{-0.2mm}\int_0^\infty F^2(m)\hspace{0.2mm}dm,
  \label{eq:Virial}
  \end{equation}
  where the dimensionless coefficients $\varw_i$ are given in equation (3.12) of C21. 
  This expression is used in
    Section \ref{sec:Models} for a check of the homoeoidal expansion approach.

  The idea behind the homoeoidal expansion is to expand equations \eqref{eq:rho(x)} 
  an \eqref{eq:Chandra} up to a prescribed order for vanishing flattening parameters 
  $\epsilon \equiv 1-q_y$ and $\eta \equiv 1-q_z$, so that the spherical case is obtained 
  when $\epsilon=0$ and $\eta=0$. From the linearity of the Poisson equation, it follows 
  that at any expansion order in the flattening, the resulting truncated density-potential 
  pairs satisfy Poisson's equation. For example, at the linear order, the expansion of the 
  density in equation \eqref{eq:rho(x)} reads  
  \begin{equation}
        \frac{\rho(\xv)}{\rhon}=
        \vrtz(s)
        +(\epsilon +\eta)\!\hspace{0.5mm}\vrtu(s)
        +\big(\epsilon\tilde{y}^2\hspace{-0.5mm}+\eta \ztil^2\big)\!\hspace{0.4mm}\vrtd(s)
        \hspace{0.4mm},
  \label{eq:rho(m)_exp}
  \end{equation}
  where $s \equiv r/a=\sqrt{\hspace{0.2mm}\tx^2+\ty^2+\tz^2\hspace{0.4mm}}$ is the dimensionless
  spherical radius, and the three dimensionless spherically symmetric components are 
  \begin{equation}
        \vrtz(s)=\vrtu(s)=\rhotil(s)\hspace{0.3mm}, 
        \qquad\quad
        \vrtd(s)=\frac{1}{s}\frac{d\rhotil(s)}{ds}.
  \end{equation}
  The corresponding expansion of the potential reads
  \begin{equation}
        \frac{\Psi(\xv)}{\Psin}=
        \tpsiz(s)
        +(\epsilon +\eta)\hspace{0.2mm}\tpsiu(s)
        +\big(\epsilon {\tilde y}^2\hspace{-0.5mm}+\eta \ztil^2\big)\tpsid(s)
        \hspace{0.4mm},
  \label{eq:Psi(x)_exp}
  \end{equation}
  where the dimensionless spherically symmetric components are
  \begin{equation}
        \tpsii(s)=
        \begin{dcases}
               \hspace{0.2mm}\frac{1}{s}\int_0^s\rhotil(m)\hspace{0.2mm}m^2dm\hspace{0.3mm}
               +\int_s^{\infty}\rhotil(m)\hspace{0.2mm}m\hspace{0.2mm}dm\hspace{0.3mm}, 
               \\[3pt]
               \hspace{0.2mm}\frac{1}{3s^3}\int_0^s\rhotil(m)\hspace{0.2mm}m^4dm\hspace{0.3mm}
               +\hspace{0.3mm}\frac{1}{3}\int_s^{\infty} \rhotil(m)\hspace{0.2mm}m\hspace{0.2mm}dm\hspace{0.3mm},
               \\[3pt]
               \hspace{0.2mm}-\hspace{0.4mm}\frac{1}{s^5} \int_0^s \rhotil(m)\hspace{0.2mm}m^4dm\hspace{0.1mm}.
        \end{dcases}
  \label{eq:Psi_i_012}
  \end{equation}
  To be physically acceptable, the truncated expanded density must be nowhere negative. 
  This requirement sets an upper limit on the possible values of $\epsilon$ and $\eta$, 
  as a function of the specific density profile adopted; for a monotonically decreasing 
  $\tilde\rho (s)$, the positivity of the right hand side of equation \eqref{eq:rho(m)_exp} 
  is assured provided that
  \begin{equation}
        \frac{1+\epsilon+\eta}{\eta} \geq A_M
        \equiv
        \sup_{s\geq 0}\hspace{0.2mm}\left|\frac{d\ln\tilde\rho(s)}{d\ln s}\right|,
  \label{eq:pos_cond_general}
  \end{equation}
  (CB05; see also exercise 2.11 in C21).
  
%%%%%%%%%%%%%%%%%%%%%%%%%%%%%%%%%%%%%%%%%%%%%%%%%%%%%%%%%%%%%%%%%%%%%%%
  \subsection{Axisymmetric oblate systems} \label{sec:Rz}
%%%%%%%%%%%%%%%%%%%%%%%%%%%%%%%%%%%%%%%%%%%%%%%%%%%%%%%%%%%%%%%%%%%%%%%  
  
  In this work we shall focus on stellar-dynamical models slightly departing 
  from spherical symmetry, 
  being qualitatively oblate: this is obtained by setting in the previous formulae 
  $\epsilon=0$, so that $\eta$ is the only flattening parameter. Moreover, as in some of 
  the following computations it is convenient to work with the truncated density-potential 
  pair explicit in $R$, we recast equations \eqref{eq:rho(m)_exp}-\eqref{eq:Psi(x)_exp} by 
  using the identity $z^2=r^2-R^2$, so that finally 
  \begin{equation}
        \begin{cases}
              \displaystyle \hspace{0.62mm}\frac{\rho(R,z)}{\rhon}=
                            \rtz(s)
                            +\eta\hspace{0.2mm}\rtu(s)
                            +\eta\hspace{0.2mm}\Rtil^2\hspace{-0.1mm}\rtd(s)\hspace{0.3mm},
              \\[12pt]
              \displaystyle \hspace{0.0mm}\frac{\Psi(R,z)}{\Psin}=
                            \tPsiz(s)
                            +\eta\hspace{0.4mm}\tPsiu(s)
                            +\eta\hspace{0.2mm}\Rtil^2\hspace{0.25mm}\tPsid(s)\hspace{0.35mm},
        \end{cases}
  \label{eq:rho_Psi}
  \end{equation}
  where $\Rtil \equiv R/a$, and the new dimensionless radial functions $\rho_i$ and 
  $\Psi_i$ are related to the dimensionless radial functions $\varrho_i$ and $\psi_i$ as
  \begin{equation}
        \begin{cases}
              \hspace{0.48mm}\rtz=\vrtz\hspace{0.3mm},
              \hspace{4mm}
              \rtu=\vrtu+s^2\vrtd\hspace{0.3mm},
              \hspace{4mm} 
              \rtd=-\hspace{0.5mm}\vrtd\hspace{0.3mm},
              \\[3pt]
              \tPsiz=\tpsiz\hspace{0.3mm},
              \hspace{3.45mm} 
              \tPsiu=\tpsiu+s^2\hspace{0.2mm}\tpsid\hspace{0.3mm},
              \hspace{3.12mm} 
              \tPsid=-\hspace{0.7mm}\tpsid\hspace{0.1mm}.
        \end{cases}
  \label{eq:rhoi_Psii}
  \end{equation}  
  
%%%%%%%%%%%%%%%%%%%%%%%%%%%%%%%%%%%%%%%%%%%%%%%%%%%%%%%%%%%%%%%%%%%%%%%%
  \section{The Jeans equations}\label{sec:JEs}
%%%%%%%%%%%%%%%%%%%%%%%%%%%%%%%%%%%%%%%%%%%%%%%%%%%%%%%%%%%%%%%%%%%%%%%%
  
  We assume that the axisymmetric model density $\rho(R,z)$, with potential 
  $\Psi(R,z)$, 
  is supported by a two-integral phase-space distribution function $f({\cal E},J_z)$, where 
  ${\cal E}$ and $J_z$ are the energy and orbital angular momentum $z$-component of stars 
  (per unit mass), respectively. We indicate with $\vR$, $\vphi$ and $\vz$ the velocity, 
  and with a bar over a quantity its average value over the velocity space. As is well known 
  (e.g. BT08, C21), for such a system: 
  (1) $\overline{\vR\vz}=\overline{\vR\vphi}=\overline{\vphi\vz}=0$; 
  (2) the only possible non-zero streaming motion is in the azimuthal direction, i.e. $\vphib$; 
  (3) at each point in the system, $\sigma_R^2=\sigma_z^2\equiv\sigma^2$.
  The Jeans equations reduce to 
  \begin{equation}
      \begin{cases}
        \displaystyle\frac{\partial\rho\hspace{0.15mm}\sigma^2}{\partial z}=\rho\hspace{0.2mm}\frac{\partial\PsiT}{\partial z},
        \\[8pt]
        \displaystyle\frac{\partial\rho\hspace{0.15mm}\sigma^2}{\partial R}-\frac{\rho\hspace{0.15mm}\Delta}{R}=\rho\hspace{0.2mm}\frac{\partial\PsiT}{\partial R},
        \qquad
        \Delta \equiv \overline{\vphi^2}-\sigma^2,
      \end{cases}
  \label{eq:Jeans}
  \end{equation}
  where $\PsiT$ is the total relative potential that can take into account the effects of other 
  density components, such as a dark matter halo and/or a central
  BH. Given the consolidated belief in a widespread presence of BHs at the center 
  of stellar systems as galaxies of various types
    (e.g., Kormendy \& Ho 2013), in order to make this work more general we consider
  \begin{equation}
        \PsiT(R,z)=\Psi(R,z)+\frac{G\Mbh}{r}\hspace{0.1mm},
  \label{eq:PsiT}
  \end{equation}
  obtained by adding to $\Psi$ the contribution of a central BH of mass $\Mbh$ 
  (of course the contribution is that of a central point mass;
  for conciseness, here and in the following, we refer in short to the BH instead of referring to its modeling through a point-mass).
  To split $\overline{\vphi^2}$ into its ordered ($\vphib$) and dispersion ($\sigphi$)  
  components, we adopt the phenomenological Satoh (1980) $k$-decomposition:
  \begin{equation} 
        \vphib=k\hspace{0.2mm}\sqrt{\Delta\hspace{0.3mm}}\hspace{0.4mm},
        \qquad
        \sigphi^2=\sigma^2+(1-k^2)\hspace{0.1mm}\Delta;
        \label{eq:Satoh}
  \end{equation} 
  $k=1$ corresponds to the isotropic rotator, while for $k=0$ no net rotation is present. Usually, $k$ 
  is assumed constant with $|k| \leq 1$; however, more general decompositions are possible, 
  with $k$ depending on $R$ and $z$ (see e.g. Ciotti \& Pellegrini 1996). It is important 
  to note that the possibility of using the Satoh decomposition depends on the positivity 
  of $\Delta$, a condition that can be violated in some proposed models, such as those with prolate 
  densities (see Section 13.3.2 in C21 and exercises 13.28 and 13.29 therein).

%%%%%%%%%%%%%%%%%%%%%%%%%%%%%%%%%%%%%%%%%%%%%%%%%%%%%%%%%%%%%%%%%%%%
  \subsection{The vertical Jeans equation}\label{subsec:vert}
%%%%%%%%%%%%%%%%%%%%%%%%%%%%%%%%%%%%%%%%%%%%%%%%%%%%%%%%%%%%%%%%%%%%

  The velocity dispersion $\sigma$ is obtained by integrating the first of the Jeans 
  equations \eqref{eq:Jeans} at fixed $R$, and imposing the boundary condition of a 
  vanishing `pressure' $\rho\hspace{0.15mm}\sigma^2=0$ for $z \to \infty$, so that
  \begin{equation}
        \rho\hspace{0.15mm}\sigma^2
        =-\!\hspace{0.2mm}\int_z^\infty \!\rho\hspace{0.3mm}\frac{\partial\PsiT}{\partial z'}\hspace{0.3mm}dz'
        =-\!\hspace{0.2mm}\int_r^\infty \!\rho\hspace{0.3mm}\frac{\partial\PsiT}{\partial r'}\hspace{0.3mm}dr'.
  \label{eq:Jeans_vert_sol}
  \end{equation}
  The second expression, where $r'=\sqrt{R^2+z'^2}$, is particularly useful when adopting 
  the explicit\hspace{0.2mm}-$R$ formulation in equation \eqref{eq:rho_Psi}: since the 
  integration is performed at fixed $R$, and the functions $\rho_i$ and $\Psi_i$ are 
  spherically symmetric, as well as the potential of the central BH, the expression of 
  $\rho\hspace{0.15mm}\sigma^2$ reduces to the evaluation of a number of integrals over the 
  spherical radius.
  
  The general considerations in the Introduction can now be made quantitative. If the 
  expanded density-potential pair is interpreted as the first order expansion of the 
  ellipsoidal parent model, {\it only zero and first order terms in the flattening must 
  be retained in equation \eqref{eq:Jeans_vert_sol}}, obtaining the so-called $\eta$-linear 
  case; in the second interpretation (the $\eta$-quadratic case), instead, the Jeans 
  equations will contain {\it up to quadratic terms in the flattening}. It is important 
  to stress that $\eta$-quadratic models are {\it not} the quadratic expansion of the 
  solutions of the Jeans equations of the original ellipsoidal system  (the full solutions), 
  since two quadratic 
  terms in the flattening are missing when using equation \eqref{eq:rho_Psi} in equation
  \eqref{eq:Jeans_vert_sol}. The quadratic expansion of the full solutions would be obtained 
  by expanding the density-potential pair up to the quadratic order included, so that in 
  equation \eqref{eq:rho_Psi} also the terms $\rho_3$ and $\Psi_3$ appear. Then, one should
  truncate the solution of equation \eqref{eq:Jeans_vert_sol} up to the quadratic order in 
  $\eta$ discarding the cubic and quartic terms in $\eta$. In practice, the quadratic expansion 
  of the full solution is given by the $\eta$-quadratic solution plus the two terms involving 
  the integrals of $\rho_0$ and $\partial\Psi_3/\partial z$, and $\rho_3$ and 
  $\partial\Psi_0/\partial z$. In Section 4.1 we show some non-trivial effects of the ``missing 
  terms''  in the $\eta $-quadratic solution, by comparing the (numerical) full solution with 
  those of the $\eta$-linear and $\eta$-quadratic models.

%%%%%%%%%%%%%%%%%%%%%%%%%%%%%%%%%%%%%%%%%%%%%%%%%%%%%%%%%%%%%%%%%%%%%%%%%
  \subsection{The radial Jeans equation}\label{sec:radial}
%%%%%%%%%%%%%%%%%%%%%%%%%%%%%%%%%%%%%%%%%%%%%%%%%%%%%%%%%%%%%%%%%%%%%%%%%
   
  Once the vertical Jeans equation is solved, no further integration would be required 
  since $\Delta$ can be evaluated from the second of the Jeans equations \eqref{eq:Jeans} as 
  a derivative,
  \begin{equation}
     \begin{aligned}
        \frac{\rho\hspace{0.15mm}\Delta}{R}
        &=\frac{\partial\rho\hspace{0.15mm}\sigma^2}{\partial R}
        -\rho\hspace{0.15mm}\frac{\partial\PsiT}{\partial R}
        \\[8pt]
        &=\int_z^{\infty}\!
               \left(
               \frac{\partial\PsiT}{\partial R}\frac{\partial\rho}{\partial z'}-
               \frac{\partial\PsiT}{\partial z'}\frac{\partial\rho}{\partial R}
               \right)\!dz'
               \equiv \hspace{0.1mm}[\hspace{0.2mm}\PsiT,\rho]\hspace{0.4mm}.
     \end{aligned}
  \label{eq:commutator}
  \end{equation}  
  The second expression above, in terms of a commutator between potential and density is 
  however to be preferred, and it has been already discussed in hydrodynamical and 
  stellar-dynamical modelling
  applications (e.g. Rosseland 1926; Hunter 1977; Waxman 1978; Barnab\`e et al. 2006; 
  CMPZ21; see also C21 and references therein). The use of the commutator reveals immediately 
  properties of $\Delta$ that in the brute-force approach of derivation are buried in the algebra.
  For example, it is immediate to show that, for any pair of spherically symmetric functions, 
  the commutator vanishes; therefore, it follows that in the Satoh decomposition spherical models  
  are necessarily isotropic (i.e. they cannot rotate), independently of the value of $k$. 
  The commutator in equation \eqref{eq:commutator} obeys several interesting and useful rules. 
  As an example, relevant for the homoeoidal expansion, is the identity
  \begin{equation}
     \begin{aligned}
        [f(R\hspace{0.25mm})\hspace{0.2mm}u(r)\hspace{0.25mm},g(R\hspace{0.25mm})\hspace{0.2mm}\varv(r)]
        =\hspace{0.8mm}&\frac{df(R\hspace{0.25mm})}{dR}\hspace{0.4mm}g(R\hspace{0.25mm})\int_r^{\infty}u(x)\hspace{0.3mm}\frac{d\varv(x)}{dx}\hspace{0.3mm}dx
        \\[6pt]
        &-f(R\hspace{0.25mm})\hspace{0.4mm}\frac{dg(R\hspace{0.25mm})}{dR}\int_r^{\infty}\frac{du(x)}{dx}\hspace{0.4mm}\varv(x)\hspace{0.3mm}dx
        \hspace{0.2mm}.
     \end{aligned}
  \label{eq:commutator_new}
  \end{equation}
  
  %%%%%%%%%%%%%%%%%%%%%%%%%%%%%%%%%%%%%%%%%%%%%%%%%%%%%%%%%%%%%%%%%%%%%%%%%%%%%%
  \begin{figure*}
        \centering
        \includegraphics[width=0.8\linewidth]{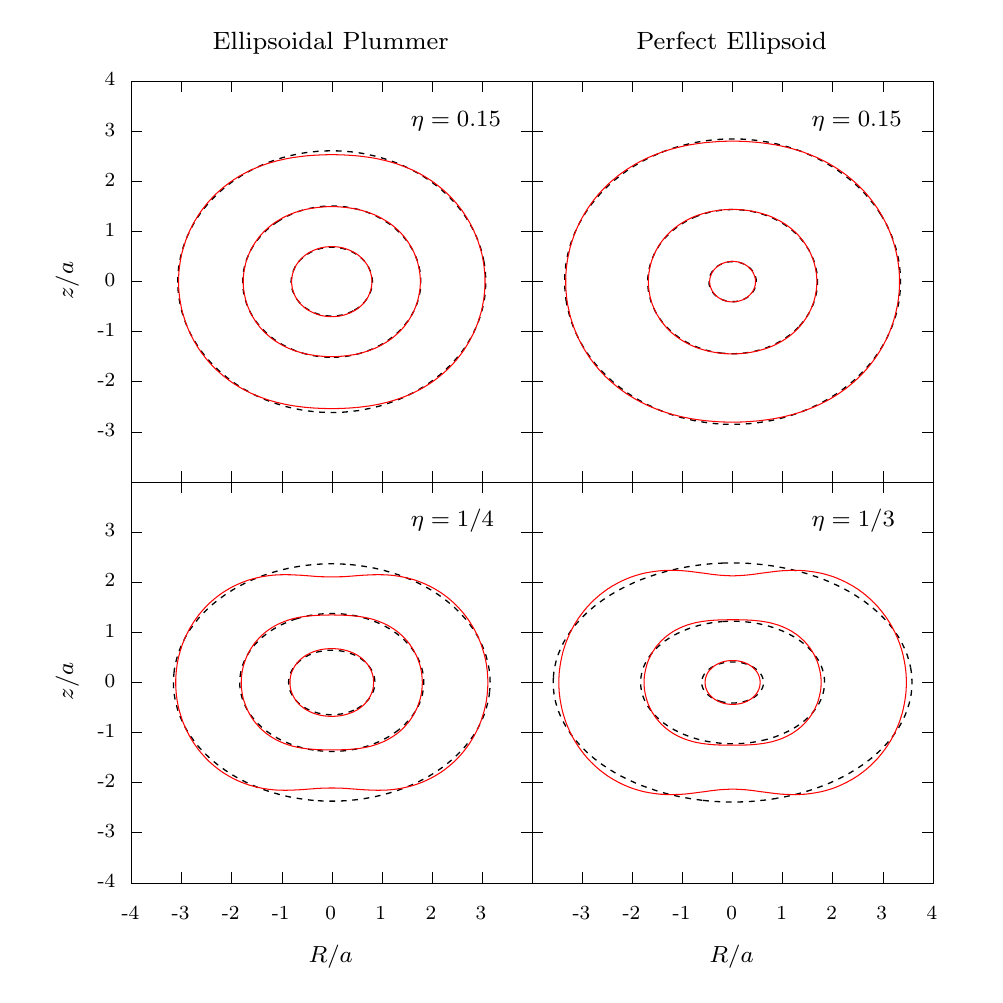}
        \vspace{-5mm}
        \caption{Isodensity contours, normalised to $\rhon$, for the 
                 P11 (left) and ZL85 (right) models. Dashed lines refer to the ellipsoidal 
                 (original) model, while solid red lines to the $\eta$-linear expansion of the
                 density, as given in equation \eqref{eq:rho_Psi}. Contours correspond to 
                 values of $1$, $10^{-1}$, and $10^{-2}$ from inside to outside. The bottom 
                 panels show the case of the critical flattenings $\eta=1/4$ for P11 models, 
                 and $\eta=1/3$ for ZL85 ones: for larger values of $\eta$ the truncated density 
                 in equation \eqref{eq:rho_Psi} would be negative near the $z$-axis (see 
                 equation \ref{eq:pos_cond_general}). The outermost expanded contours differ 
                 most from the elliptical shape as $\eta$ increases.}
  \label{fig:fig1}
  \end{figure*}
  %%%%%%%%%%%%%%%%%%%%%%%%%%%%%%%%%%%%%%%%%%%%%%%%%%%%%%%%%%%%%%%%%%%%%%%%%%%%%%
  %%%%%%%%%%%%%%%%%%%%%%%%%%%%%%%%%%%%%%%%%%%%%%%%%%%%%%%%%%%%%%%%%%%%%%%%%%%%%%
  \begin{figure*}
        %\centering
        \includegraphics[width=0.45\linewidth]{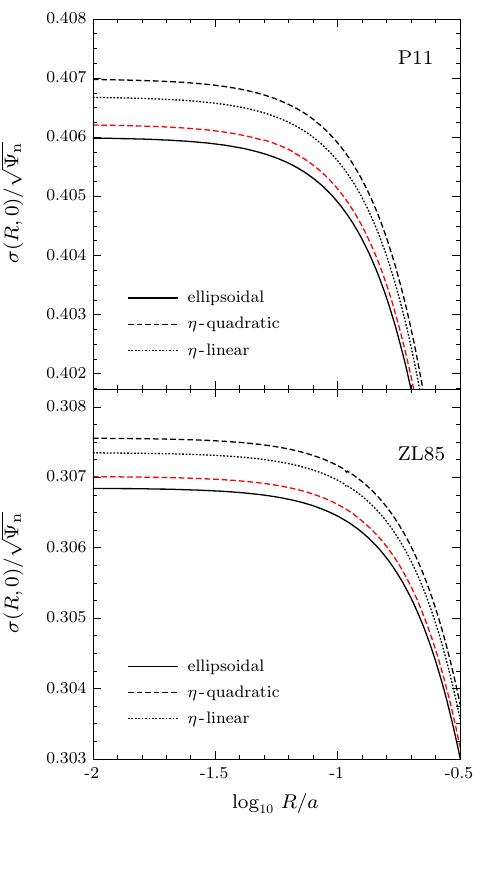}
        \hspace{3mm}
        \includegraphics[width=0.45\linewidth]{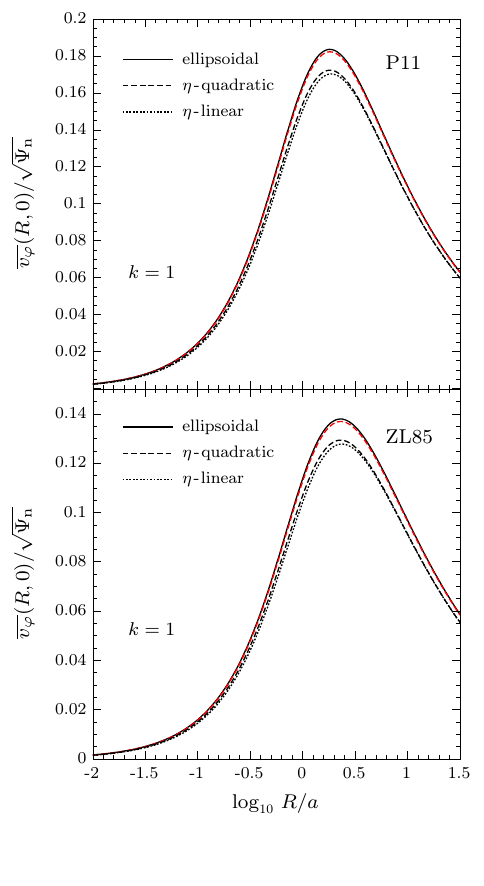}
        \vspace{-11.5mm}
        \caption{Left: radial profile of the velocity dispersion $\sigma$ in the equatorial 
                 plane ($z=0$) for the models discussed in Section \ref{sec:Models}, with 
                 $\eta=0.1$, and without a central BH ($\mu=0$). The homoeoidal approximation 
                 reproduces remarkably well the full solution (solid line), over the displayed 
                 radial range of $0.01<R/a\lesssim 0.32$. The values of $\sigma$ for the 
                 $\eta$-linear models overestimate the full solution, and retaining the quadratic 
                 terms in the flattening further increases this overestimate. Right: radial trend 
                 of $\vphib$ on the equatorial plane, for the same models on the left, and
                 with the same meaning of the line-type; the case of the isotropic rotator is 
                 shown. The values of $\vphib$ for the $\eta$-linear and $\eta$-quadratic 
                 models underestimate those of the full solution. In all panels, the red dashed 
                 line shows the expansion up to the quadratic order in $\eta $ of the full solution 
                 (see equation \ref{eq:rhosig2_trueQUAD}, and Section \ref{sec:linVSquad}).}
  \label{fig:fig2}
  \end{figure*}
  %%%%%%%%%%%%%%%%%%%%%%%%%%%%%%%%%%%%%%%%%%%%%%%%%%%%%%%%%%%%%%%%%%%%%%%%%%%%%%

  %%%%%%%%%%%%%%%%%%%%%%%%%%%%%%%%%%%%%%%%%%%%%%%%%%%%%%%%%%%%%%%%%%%%%%%%%
  \subsection{The $\eta$-quadratic solution of the Jeans equations}\label{sec:Solution}
  %%%%%%%%%%%%%%%%%%%%%%%%%%%%%%%%%%%%%%%%%%%%%%%%%%%%%%%%%%%%%%%%%%%%%%%%%

  In this Section we present the general $\eta$-quadratic solution for the homoeoidal 
  expansion of a model with a central BH of mass $\Mbh=\mu M$; quite obviously, the 
  $\eta$-linear solution is just obtained by ignoring the quadratic terms. As usual, 
  we split the velocity dispersion into the contribution $\sigs$, due to the 
  potential of the model for the stellar component, and $\sigBH$, the one 
  due to the potential of the BH: 
  $\sigma^2=\sigs^2+\sigBH^2$. By inserting the expansions \eqref{eq:rho_Psi} in 
  equation \eqref{eq:Jeans_vert_sol}, with $\PsiT$ given by equation \eqref{eq:PsiT}, 
  some algebra shows that
  \begin{equation}
         \begin{aligned}
               \frac{\rho\hspace{0.15mm}\sigs^2}{\rhon\Psin} 
               =\hspace{0.8mm}&\frac{\Hzz(s)}{2}
               +\eta\hspace{0.2mm}
                    \big[
                    \hspace{0.15mm}\Hzu(s)
                    +\Rtil^2\Hzd(s)\hspace{0.1mm}
                    \big]
               \\[4pt]
               &+\eta^2\hspace{-0.4mm}
                    \left[\hspace{0.2mm}\frac{\Huu(s)}{2}
                    +\Rtil^2\Hud(s)
                    +\Rtil^4\hspace{0.2mm}\frac{\Hdd(s)}{2}\hspace{0.1mm}
                    \right]\hspace{-0.4mm},
         \end{aligned}
  \label{eq:Hij}
  \end{equation}
  and
  \begin{equation}
        \frac{\rho\hspace{0.15mm}\sigBH^2}{\rhon\Psin}=\mu\hspace{0.1mm}
                            \big[\hspace{0.1mm}
                            Y_0(s)
                            +\eta\hspace{0.2mm}Y_1(s)
                            +\eta\hspace{0.2mm}\Rtil^2\hspace{-0.1mm}Y_2(s)\hspace{0.1mm}
                            \big]\hspace{0.2mm},
  \label{eq:Xi}
  \end{equation}
  where, for $i,j=0,1,2$, the dimensionless functions $\Hij(s) \equiv \Xij(s)+X_{ji}(s)$ 
  and $Y_i(s)$ are given by
  \begin{equation}
        \Xij(s)\equiv
        -\int_s^\infty\rho_i(t)\hspace{0.2mm}\frac{d\Psi_j(t)}{dt}\hspace{0.25mm}dt\hspace{0.2mm},
        \qquad
        Y_i(s)\equiv
        \int_s^\infty\frac{\rho_i(t)}{t^2}\hspace{0.2mm}dt\hspace{0.2mm},
        \label{eq:X_Y}
  \end{equation}
  and $\rho_i$ and $\Psi_i$ are given in equation \eqref{eq:rhoi_Psii}; notice that 
  $\Hij=H_{ji}$.

  We then evaluate $\Delta=\Dels+\DelBH$ from equation \eqref{eq:commutator}, where 
  $\Dels$ and $\DelBH$ are the contributions due to the potential of the stars and the
  one due to the potential of the central BH, respectively. From equations 
  \eqref{eq:commutator}, \eqref{eq:rho_Psi}, and \eqref{eq:commutator_new}, we obtain
  \begin{equation}
        \frac{\rho\hspace{0.15mm}\Dels}{\rhon\Psin}=
        2\hspace{0.4mm}\eta\hspace{0.2mm}\Rtil^2 
                    \big[
                    \Zzd(s)+\eta\hspace{0.2mm}\Zud(s)+\eta\hspace{0.2mm}\Rtil^2\Zdd(s)
                    \big]\hspace{-0.1mm},
  \label{eq:rhos_Deltag}
  \end{equation}   
  and
  \begin{equation}
        \frac{\rho\hspace{0.15mm}\DelBH}{\rhon\Psin}= 
        2\hspace{0.2mm}\mu\hspace{0.3mm}\eta\Rtil^2 Y_2(s)\hspace{0.4mm},
  \label{eq:rhos_DeltaBH}
  \end{equation} 
  where $\Zij(s) \equiv \Hij(s)-\rho_i(s)\hspace{0.2mm}\Psi_j(s)$. As expected, 
  $\Dels$ and $\DelBH$ vanish for $\eta=0$; moreover, $\sigBH^2$ and $\DelBH$ depend 
  linearly on $\eta$ also in the $\eta$-quadratic interpretation.
  
  A comment is in order here, about the fact that in the $\eta$-linear and 
  $\eta$-quadratic frameworks the expressions above refer to the products 
  $\rho\hspace{0.15mm}\sigma^2$ and $\rho\hspace{0.15mm}\Delta$, and {\it not} to the 
  purely kinematical fields $\sigma^2$ and $\Delta$. In order to obtain $\sigma^2$ and 
  $\Delta$, one must divide $\rho\hspace{0.15mm}\sigma^2$ and $\rho\hspace{0.15mm}\Delta$ 
  by the density in equation \eqref{eq:rho_Psi}, i.e., by a linear function in $\eta$.
  In the $\eta$-linear case one should then expand the fraction up to linear terms in 
  $\eta$; in the $\eta$-quadratic interpretation, instead, where the density-potential 
  pair in equation \eqref{eq:rho_Psi} is considered a model by itself, one should not 
  expand, so that the purely kinematical fields are {\it not} polynomial functions of 
  the flattening. Of course, in the limit of small flattening, even in the $\eta$-quadratic
  interpretation, the kinematical fields can be expanded up to $\eta^2$ terms included, 
  thus obtaining more manageable expressions. 
  
  %%%%%%%%%%%%%%%%%%%%%%%%%%%%%%%%%%%%%%%%%%%%%%%%%%%%%%%%%%%%%%%%%%%%%%%
  \section{The Models}\label{sec:Models}
  %%%%%%%%%%%%%%%%%%%%%%%%%%%%%%%%%%%%%%%%%%%%%%%%%%%%%%%%%%%%%%%%%%%%%%%

  We now consider axisymmetric systems of density distribution $\rho(R,z)$, total 
  mass $M$, scale length $a$, and axial ratio $q=1-\eta$, where $0\leq\eta<1$. From 
  equations \eqref{eq:m} and \eqref{eq:rho(x)} with $\epsilon=0$, one has:
  \begin{equation}
        \frac{\rho(R,z)}{\rhon}=\frac{\rhotil(m)}{1-\eta}\hspace{0.2mm},
        \qquad
        m^2\hspace{-0.3mm}=\Rtil^2+\hspace{0.2mm}\frac{\ztil^2}{(1-\eta)^2}\hspace{0.2mm}.
        \label{eq:rho(R,z)}
  \end{equation}
  In particular, we solve in closed form the $\eta$-quadratic Jeans equations for two 
  ellipsoidal models: the ellipsoidal generalization of the Plummer model (P11), and the 
  Perfect Ellipsoid (ZL85). The $\eta$-linear cases are immediately obtained by neglecting 
  the $\eta^2$ terms. The dimensionless densities of the two models are given respectively by
  \begin{equation}
        \rhotil(m)=\
        \begin{dcases}
              \hspace{0.05cm}\frac{3}{(1+m^2)^{5/2}}\hspace{0.1mm}, 
              %\hspace{0.6cm} (\Muu)\hspace{0.3mm},
              \\[4pt]
              \hspace{0.05cm}\frac{4}{\upi\hspace{0.1mm}(1+m^2)^2}\hspace{0.1mm}, 
              %\hspace{0.652cm} (\Moc)\hspace{0.3mm},
        \end{dcases}
  \label{eq:rhotil_M11M85}
  \end{equation}
  and the masses enclosed within $m$ are
  \begin{equation}
        \frac{M(m)}{M}=
        \begin{dcases}
              \hspace{0.05cm}\frac{m^3}{(1+m^2)^{3/2}}\hspace{0.1mm}, 
              %\hspace{2cm} (\Muu)\hspace{0.3mm},
              \\[4pt]
              \hspace{0.05cm}\frac{2}{\upi}\hspace{-0.2mm}\left(\hspace{0.3mm}\arctan m-\frac{m}{1+m^2}\hspace{0.2mm}\right)\hspace{-0.4mm}. 
              %\hspace{0.55cm} (\Moc)\hspace{0.1mm}.
        \end{dcases}
  \label{eq:M_M11M85}
  \end{equation}
  In the adopted notation, the circular velocity in the equatorial plane can be 
  written as
  \begin{equation}
      \frac{\vcirc^2(R\hspace{0.15mm})}{\Psin}=\frac{\Rtil^2}{2}\bigintsss_{\hspace{0.4mm}0}^{\infty}
                  \rhotil\hspace{-0.15mm}\left(\frac{\Rtil}{\sqrt{1+u\hspace{0.4mm}}}\right)
                  \hspace{-0.4mm}\frac{du}{(1+u)^2\sqrt{\hspace{0.1mm}q^2+u\hspace{0.4mm}}}
                  \hspace{0.2mm},
  \label{eq:vcirc}
  \end{equation}
  (e.g., see equation 5.62 in C21), and for P11 we obtain
  \begin{equation}
        \frac{\vcirc^2(R\hspace{0.15mm})}{\Psin}=
        \frac{p^2\hspace{0.2mm}\ellF\hspace{0.3mm}(\phi,k)+(\Rtil^2-p^2)\hspace{0.3mm}\ellE\hspace{0.3mm}(\phi,k)}{(p^2+\Rtil^2)^{3/2}}
        -\frac{q\hspace{0.1mm}\Rtil^2}{(p^2+\Rtil^2)(1+\Rtil^2)^{3/2}}\hspace{0.1mm},
  \label{eq:vc_true_P11}
  \end{equation} 
  where 
  \begin{equation}
        \phi \equiv \arcsin\sqrt{\hspace{0.25mm}\frac{p^2+\Rtil^2}{1+\Rtil^2}\hspace{0.1mm}}\hspace{0.3mm},
        \quad
        k \equiv \frac{\Rtil}{\sqrt{p^2+\Rtil^2}}\hspace{0.25mm},
        \quad
        p^2 \equiv 1-q^2,
  \label{eq:phi_p_q}
  \end{equation} 
  and $\ellF\hspace{0.3mm}(\phi,k)$ and $\ellE\hspace{0.3mm}(\phi,k)$ are the Legendre 
  elliptic integrals  of first 
  and second kind in trigonometric form (e.g. Gradshteyn \& Ryzhik 2007). For ZL85 we have:
  \begin{equation}
        \frac{\vcirc^2(R\hspace{0.15mm})}{\Psin}=
        \frac{2\Rtil^2}{\upi\hspace{0.1mm}(p^2+\Rtil^2)}\hspace{-0.1mm}
        \left(
        \hspace{0.1mm}\frac{\phi}{\sqrt{p^2+\Rtil^2}}-\frac{q}{1+\Rtil^2}\hspace{0.1mm}
        \right)\hspace{-0.4mm},
  \label{eq:vc_true_ZL85}
  \end{equation}   
  where $p$ and $\phi$ are given in equation \eqref{eq:phi_p_q}. Equations 
  \eqref{eq:vc_true_P11} and \eqref{eq:vc_true_ZL85} are exact for any finite value of $\eta$; 
  with some work, they can be expanded to any desired order in $\eta$, thus providing a check  
  for the homoeoidal expansion, where
  \begin{equation}
        \frac{\vcirc^2(R\hspace{0.15mm})}{\Psin}=
        \tvcz(\Rtil\hspace{0.2mm})
        +\eta\hspace{0.2mm}\tvcu(\Rtil\hspace{0.2mm})\hspace{0.2mm},
        \qquad
        \varv_i^2(\Rtil\hspace{0.2mm}) \equiv 
        -\hspace{0.4mm}\Rtil\hspace{0.5mm}\frac{d\psi_i(\Rtil\hspace{0.2mm})}{d\Rtil}\hspace{0.2mm}.
  \label{eq:vcirc_exp}
  \end{equation}
  We verified that the linear expansion of equations \eqref{eq:vc_true_P11} and 
  \eqref{eq:vc_true_ZL85} are in perfect agreement with equation \eqref{eq:vcirc_exp}, where 
  $\varv_0$ and $\varv_1$ are given in the Appendix.
  
  For an axisymmetric model with a central BH, the full solution of equations
  \eqref{eq:Jeans_vert_sol} and \eqref{eq:commutator} can be recast in integral form
  by exploiting the assumed homoeoidal structure. For the contributions 
  of the stellar-dynamical models 
  we have 
  \begin{equation}
        \frac{\rho\hspace{0.15mm}\sigs^2}{\rhon\Psin}=
        \frac{1}{2q}
        \bigintsss_{\hspace{0.4mm}\ztil}^\infty \hspace{-0.8mm}\rhotil(m')\hspace{0.3mm}\ztil'd\ztil'\hspace{-0.1mm}
        \hspace{-0.8mm}
        \bigintsss_{\hspace{0.4mm}0}^\infty \hspace{-0.4mm}\frac{\rhotil(m_u)\hspace{0.2mm}du}{(1+u)(q^2+u)^{3/2}}\hspace{0.2mm},
        \label{eq:rho_sigs2_true}
  \end{equation}
  and
  \begin{equation}
        \frac{\rho\hspace{0.15mm}\Dels}{\rhon\Psin}=
        \frac{p^2\Rtil^2}{2q^3}\hspace{-0.4mm}
        \bigintsss_{\hspace{0.4mm}\ztil}^\infty \hspace{0.4mm}\left|\frac{d\rhotil(m')}{dm'}\hspace{0.1mm}\right|\frac{\ztil'd\ztil'}{m'}\hspace{-0.6mm}
        \bigintsss_{\hspace{0.4mm}0}^\infty \hspace{-0.4mm}\frac{\rhotil(m_u)\hspace{0.2mm}u\hspace{0.1mm}du}{(1+u)^2(q^2+u)^{3/2}},
        \label{eq:rho_Dels_true}
  \end{equation}
  where $p^2=1-q^2$, $m_u=m({\bf x},u)$ is given by equation \eqref{eq:mu},
  $m'=\sqrt{\Rtil^2+\ztil'^2/q^2}$, and the two integrals are evaluated at fixed
  $\Rtil$ (e.g., see exercise 13.29 in C21). The contributions due to a central BH of mass 
  $\Mbh=\mu M$ are instead given by
  \begin{equation}
        \frac{\rho\hspace{0.15mm}\sigBH^2}{\rhon\Psin}=
        \frac{\mu}{q}
        \bigintsss_{\hspace{0.4mm}\ztil}^\infty \hspace{-0.2mm}\rhotil(m')\hspace{0.25mm}\frac{\ztil'd\ztil'}{s'^{\hspace{0.15mm}3}},
  \end{equation}
  and
  \begin{equation}
        \frac{\rho\hspace{0.15mm}\DelBH}{\rhon\Psin}=
        \frac{\mu\hspace{0.1mm}p^2\Rtil^2}{q^3}\hspace{-0.4mm}
        \bigintsss_{\hspace{0.4mm}\ztil}^\infty \hspace{0.4mm}\left|\frac{d\rhotil(m')}{dm'}\hspace{0.1mm}\right|\frac{\ztil'd\ztil'}{m\hspace{0.25mm}s'^{\hspace{0.15mm}3}},
        \label{eq:rho_DelBH_true}
  \end{equation}\\
  where $s'=\sqrt{\Rtil^2+\ztil'^2}$ (e.g., see exercise 13.28 in C21).
  
  Finally, for what concerns the self-gravitational energy of the models, the integral in 
  equation \eqref{eq:Virial} evaluates to $3\upi/4$ and $4/\upi$, respectively for the P11 
  models and ZL85 models, so that, by expanding up to the second order in $\eta$
  the coefficients $\varw_i$ in equation \eqref{eq:Virial}, for the axisymmetric case
  one has
  \begin{equation}
        \frac{W}{M\Psin}\sim-\hspace{0.4mm}\frac{1}{8}\hspace{-0.2mm}
        \left(1+\frac{\eta}{3}+\frac{2\hspace{0.2mm}\eta^2}{15}\right)\hspace{-0.4mm}
        \hspace{-0.2mm}\int_0^\infty \hspace{-0.2mm} F^2(m)\hspace{0.2mm}dm\hspace{0.1mm}.
  \end{equation}
  This expression can be used as a check of the homoeoidal expansion,
  when the self-gravitational energy is computed directly from the density-potential
  pair in equation \eqref{eq:rho_Psi}, limiting to the linear terms in $\eta$.
  
  %%%%%%%%%%%%%%%%%%%%%%%%%%%%%%%%%%%%%%%%%%%%%%%%%%%%%%%%%%%%%%%%%%%%%%%%%%%%%%%%%%%%
  \begin{figure*}
        %\centering
        \includegraphics[width=0.45\linewidth]{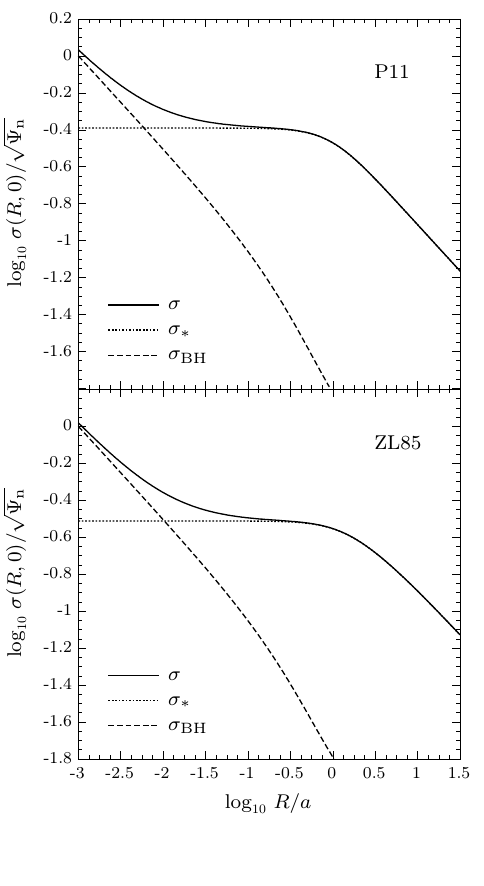}
        \hspace{3mm}
        \includegraphics[width=0.45\linewidth]{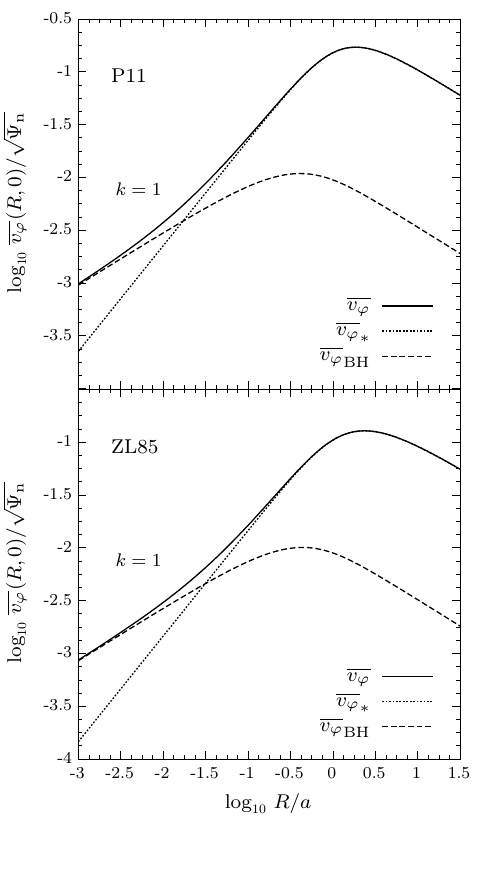}
        \vspace{-12mm}
        \caption{The $\eta$-linear modelling of an isotropic ($k=1$) system with 
                 $\eta=0.1$, and a central BH with $\mu=0.001$, for the models in Section
                 \ref{sec:Models}. Solid lines show $\sigma$ (left) and $\vphib$ (right) on 
                 the equatorial plane ($z=0$). Dotted and dashed lines show respectively the
                 contributions of the stellar-dynamical model and the BH. Due to the presence of the BH, 
                 the velocity dispersion diverges at small radii as $1/\sqrt{R}$, while the 
                 streaming velocity vanishes towards the center as $\sqrt{R}$ (see Section
                 \ref{sec:linVSquad}).}
  \label{fig:fig3}
  \end{figure*}  
  %%%%%%%%%%%%%%%%%%%%%%%%%%%%%%%%%%%%%%%%%%%%%%%%%%%%%%%%%%%%%%%%%%%%%%%%%%%%%%%%%%%%

  \subsection{Results}\label{sec:linVSquad}

  As recalled in Section \ref{sec:HomExp}, in the homoeoidal expansion there is an upper 
  limit on $\eta$ (that depends on the truncation order), so that, for $\eta$ smaller than
  the critical value, the truncated density is nowhere negative: from \eqref{eq:pos_cond_general}, 
  $\eta \leq 1/4$ for P11 models, and $\eta \leq 1/3$ for ZL85 models. Reassuringly, these 
  critical values are quite large, allowing to deal with moderately flattened stellar systems 
  such as those discussed in Section \ref{sec:NGC4372}. In general, for $\eta$ close to the 
  limit, the density tends to become negative along the $z$-axis, producing densities with a 
  `torus-like' structure, similar to the Binney logarithmic halo for potential flattening
  near the critical value (BT08), and to complex shifted models (e.g. Ciotti \& Giampieri 2007).
  In Fig. \ref{fig:fig1} we show the isodensity contours of P11 and ZL85 models, for two 
  different $\eta$ values. Black dashed lines show the original ellipsoidal models in 
  equation \eqref{eq:rhotil_M11M85}, while red solid lines show the homoeoidally expanded 
  models in equation \eqref{eq:rho_Psi}, where the explicit expressions for $\rho_i$ are  
  given in Appendix. The figure shows how well the truncated density reproduces the original
  model with $\eta=0.15$, and how a toroidal shape in the outer parts of the systems appears 
  for $\eta$ approaching the critical value. Of course, truncating the density up to the 
  quadratic order in $\eta$ increases the upper limit on the flattening, and both black and 
  red isdodensities would be almost indistinguishable also in the analogous of Fig. 
  \ref{fig:fig1} (not shown here for simplicity). 
  
  %%%%%%%%%%%%%%%%%%%%%%%%%%%%%%%%%%%%%%%%%%%%%%%%%%%%%%%%%%%%%%%%%%%%%%%%%%%%%%
  \begin{figure*}
        %\centering
        \includegraphics[width=0.35\linewidth]{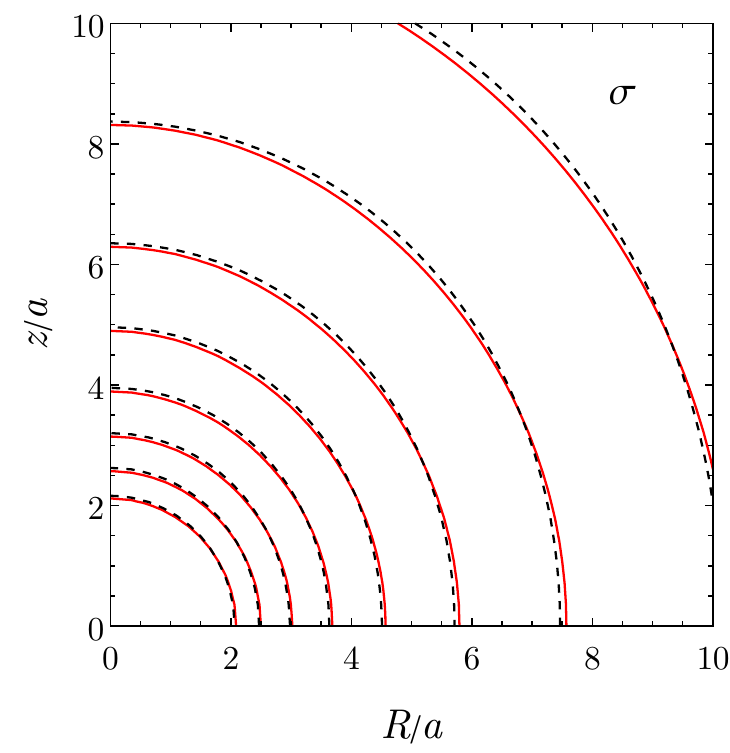}
        \hspace{-6mm}
        \includegraphics[width=0.35\linewidth]{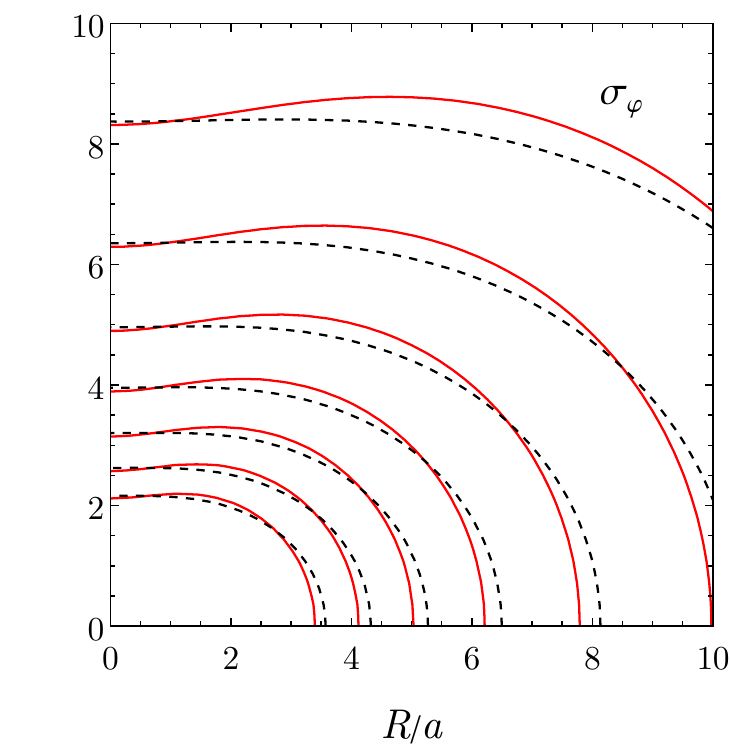}
        \hspace{-6mm}
        \includegraphics[width=0.35\linewidth]{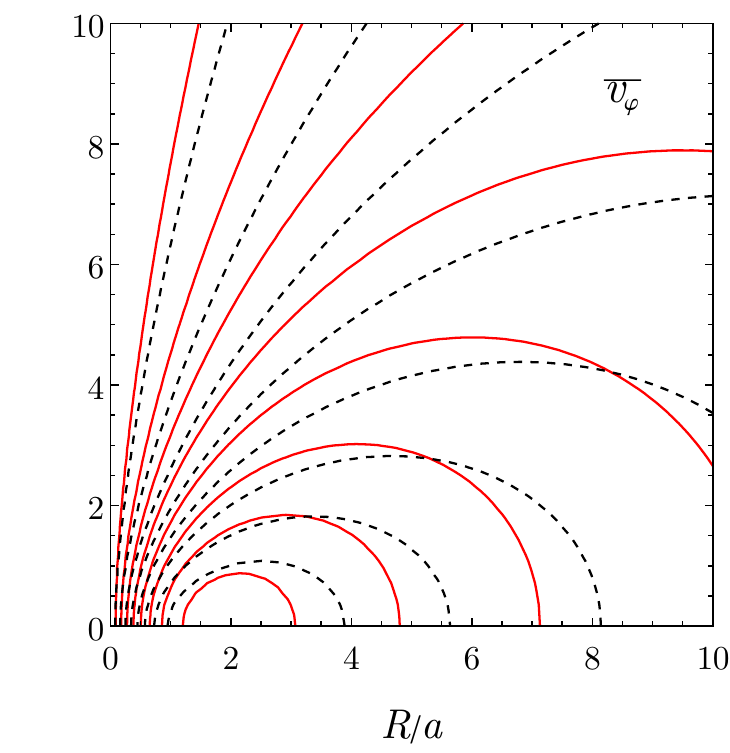}
        \vspace{-1mm}
        \caption{Maps of $\sigma$ (left), $\sigphi$ (middle, $k=0$), and $\vphib$ (right, 
                 $k=1$), in units of $\sqrt{\Psin}$, for P11 models with $\eta=0.1$ and 
                 $\mu=0$. Dashed lines show the full solution, solid red lines show the 
                 solution of the $\eta$-linear modelling. The innermost contour  
                 corresponds to values of $0.26$ for the normalized $\sigma$ and $\sigphi$, 
                 and of $0.16$ for $\vphib$. The values for the other contour lines decrease  
                 outward with steps of $0.02$.}
  \label{fig:fig4}
  \end{figure*}
  %%%%%%%%%%%%%%%%%%%%%%%%%%%%%%%%%%%%%%%%%%%%%%%%%%%%%%%%%%%%%%%%%%%%%%%%%%%%%% 
  
  In order to address the first goal of this work, i.e., to evaluate how $\eta$-linear 
  and $\eta$-quadratic solutions compare between them and with respect to the full solutions 
  for genuine ellipsoidal models, in Fig. \ref{fig:fig2} we show the velocity dispersion 
  $\sigma$, and the streaming velocity $\vphib$ in the equatorial plane, for P11 and ZL85 
  models for a flattening $\eta=0.1$, and in the isotropic rotator ($k=1$). The solutions 
  are shown in absence of the central BH for simplicity; the explicit expressions for the 
  functions $X_i$, $Y_i$, $H_{ij}$ and $Z_{ij}$, entering $\rho\hspace{0.15mm}\sigs^2$ and 
  $\rho\hspace{0.15mm}\Dels$ in equations \eqref{eq:Hij} and \eqref{eq:rhos_Deltag}, are given 
  in the Appendix. We first focus on $\sigma$. For each model, the left panels of Fig. 
  \ref{fig:fig2} show the full solution for the ellipsoidal model (black solid line), obtained 
  by solving numerically equation \eqref{eq:rho_sigs2_true}, the $\eta$-linear solution
  (dotted line), the $\eta$-quadratic solution (dashed line), and the truncated expansion of 
  the full solution up to the quadratic order terms included (red dashed line). For reference, 
  the central values of $\sigma$ in the $\eta$-quadratic solutions are
  \begin{equation}
     \frac{\sigma_0^2}{\Psin}=
     \begin{cases}
         \displaystyle \frac{2+4(67-96\ln 2)\hspace{0.2mm}\eta+3(355-512\ln 2)\hspace{0.2mm}\eta^2}{12(1+\eta)}
         \hspace{0.2mm},
         %\hspace{0.991cm} (\Muu)\hspace{0.3mm}
         \\[14pt]
         \displaystyle \frac{3(32-3\upi^2)+8(9\upi^2-88)\hspace{0.2mm}\eta+24(15\upi^2-148)\hspace{0.2mm}\eta^2}{24\upi(1+\eta)}
         \hspace{0.2mm},
         %\hspace{0.15cm} (\Moc)\hspace{0.3mm}
     \end{cases}
     \label{eq:sigma_0}
  \end{equation}
  for the P11 and ZL85 models respectively. Obviously, the $\eta$-linear case is obtained 
  by neglecting the $\eta^2$ terms at the numerators. A few general features are apparent.
  The first is the expected similarity of the full solution for the two models, due to the 
  qualitatively similar behaviour of their density distributions in the central regions. In 
  the external regions, the decline of $\sigma$ in both models goes as $R^{-1/2}$ 
  being $\rho\hspace{0.15mm}\sigs^2 \propto r^{-\hspace{0.3mm}6}$ for the P11 models, and 
  $\rho\hspace{0.15mm}\sigs^2 \propto r^{-\hspace{0.3mm}5}$ for the ZL85 ones.
  
  The second  reassuring feature is how close the full solution and those in the homoeoidal
  approximation are, over the whole radial range: the percentual differences are so small 
  (less than $0.3 \%$) to be completely negligible in all practical applications. Therefore, 
  we can conclude that the effect of quadratic $\eta$ terms is negligible, and that the 
  $\eta$-linear approximation, with its simplifications, can be safely used to model systems  
  with low flattening.
  
  We can now address an interesting result that emerges from Fig. \ref{fig:fig2}, also completing
  the reasonings introduced at the end of Section 3.3. For both models, the $\eta$-linear solution
  always overestimates the full solution (with differences decreasing for increasing $R$), and so 
  do the other approximations; however,  the $\eta$-quadratic solutions differ from the full 
  solution {\it more} than the $\eta$-linear solutions. This result might be unexpected, since a
  quadratic approximation should perform better than a linear one. 
  But it should be recalled that the $\eta$-quadratic solution is {\it not} the quadratic 
  approximation of the full solution. In fact, the dashed red lines in Fig. \ref{fig:fig2} 
  confirm that the quadratic approximation $(\rho\hspace{0.15mm}\sigs^2)_{\rm quadratic}$
  performs better than the $\eta$-linear solution $(\rho\hspace{0.15mm}\sigs^2)_{\eta-{\rm linear}}$.
  As discussed in Section \ref{sec:Solution}, the quadratic expansion of the full solution
  \begin{equation}
     (\rho\hspace{0.15mm}\sigs^2)_{\rm quadratic}\simeq
     (\rho\hspace{0.15mm}\sigs^2)_{\rm \eta-{\rm linear}}+\eta^2 Q
     \label{eq:rhosig2_trueQUAD}
  \end{equation} 
  could be computed formally starting from the homoeoidal truncation of the density-potential 
  pair to the quadratic order in $\eta$, solving the Jeans equations, and finally discarding
  all terms in flattening of order higher than quadratic. However, instead of performing such 
  laborious mathematical calculations, we computed numerically the function $Q$ in equation
  \eqref{eq:rhosig2_trueQUAD} as
  \begin{equation}
     Q \equiv
     \lim_{\eta \to 0}\frac{\rho\hspace{0.15mm}\sigs^2
     -(\rho\hspace{0.15mm}\sigs^2)_{\rm \eta-{\rm linear}}}{\eta^2}\hspace{0.2mm},
  \end{equation} 
  where $\rho\hspace{0.15mm}\sigs^2$ is the full (numerical) solution. In the formula above, 
  the numerical value of $\eta$ are reduced until convergence is reached (but maintained 
  large enough to avoid numerical fluctuations). The fact that the $\eta$-quadratic solution 
  is {\it not} the quadratic truncation of the expansion of the full solution is made
  apparent by the fact that the black dashed lines (the $\eta$-quadratic solutions) are more 
  distant from the solid line than the $\eta$-linear solution (dotted lines); this is due to 
  the missing quadratic terms, which can be shown to be collectively negative. The conclusion 
  is that, when using the homoeoidal expansion to describe an ellipsoidal system, the 
  $\eta$-linear interpretation is to be preferred to the $\eta$-quadratic solution, not only 
  for its greater simplicity, but also for its better accuracy.
  
  The right panels of Fig. \ref{fig:fig2} show the corresponding streaming velocity profile 
  $\vphib$ in the equatorial plane for the isotropic case ($k=1$); the radial range has been 
  extended to $R \simeq 32\hspace{0.4mm}a$, in order to display the whole peak present at 
  around $R \simeq 1.6\hspace{0.4mm}a$. The Satoh decomposition can be adopted for these models 
  given the positivity of $\Dels$, which is to be expected since $\Delta$ is nowhere negative 
  for an oblate  self-gravitating ellipsoid, as shown by equation \eqref{eq:rho_Dels_true}.
  Several of the comments concerning the solutions for the velocity dispersion apply also to 
  $\vphib$,
  in particular that on the almost perfect (for practical purposes)
  coincidence of the $\eta$-linear, $\eta$-quadratic, and full solutions. However, the 
  $\eta$-linear solutions are now the most discrepant with respect to the full solutions, 
  followed, in order, by the $\eta$-quadratic and the true quadratic expansion.
  
  The effect of a central BH of mass $\Mbh=10^{-3}\Ms$ on the
  $\eta$-linear solution is shown for the P11 and ZL85 models in
  Fig. \ref{fig:fig3}. In each plot, the solid line is the total, the
  dashed line is the BH contribution, and the dotted line is the
  model for the stellar component already shown in
  Figure \ref{fig:fig2}; the radial range is now extended down to
  $R=10^{-3}\hspace{0.2mm}a$ to better appreciate the dynamical
  effects of the BH. From equation \eqref{eq:M_M11M85}, the radius
  contaning the fraction $\mu=10^{-3}$ of the total mass (of the
  spherical model), that is the commonly adopted estimate for the
  dynamical radius of the BH (see Chapter 4 in BT08),
  is $R_{\rm dyn}\approx 0.1\hspace{0.2mm}a$. This value is nicely
  close to the position where the lines corresponding to the total
  $\sigma$ and $\vphib$ start to deviate from the
  stellar-dynamical model contributions\footnote{
    Alternatively (e.g., see BT08), the radius of the
      sphere of influence of the BH can be defined as the distance
      from the centre at which the circular velocity due to the BH
      equals the projected velocity dispersion, i.e.
      $R_{\rm infl}=G\Mbh/\siglos^2(R_{\rm infl})$. For our models, in
      the limit of spherical symmetry, and under the assumption of
      isotropic velocity dispersion, $R_{\rm infl}\simeq 6\mu a$, almost $16$
      times smaller than $R_{\rm dyn}$.}.  In
  particular, the BH determines an increase of $\sigma$ towards the
  centre that goes as $R^{\hspace{0.2mm}-1/2}$;
  instead, $\vphib$ still vanishes at the centre even in the presence
  of the BH. This property can be quantified with the asymptotic
  analysis of $\Dels$ and $\DelBH$ near the center: without the BH,
  the isotropic $\vphib$ decreases at small radii as $R$, whereas in
  presence of the BH it decreases as $R^{\hspace{0.2mm}1/2}$; thus,
  $\vphib$ does not diverge at the centre, as instead $\sigma$ and
  $\vcirc$ do. This is explained by noticing that, for a generic
  model density with a central profile
  $(1+m^2)^{-\alpha}$, $\DelBH \propto R^2/r$ at small radii,
  and so in the Satoh decomposition $\vphib$ vanishes towards the
  centre\footnote{ The vanishing of $\DelBH$ is {\it not} a general
    property of ellipsoidal systems with a central BH (e.g., see
    Fig. 3 in CMPZ21).}, while $\overline{\varv_\varphi^2}$ diverges
  as $\sigma^2$. Of course, when adopting a different decomposition of
  $\overline{\varv_\varphi^2}$ (such as that in equation 13.107 in
  C21; see also De Deo et al. 2024), a central cusp in $\vphib$ would
  be obtained. We conclude that special care should be used when
  interpreting the results of models used to predict the effects of a
  central BH on the streaming velocity field of the stars.
  
  The previous discussion focused on the different solutions on the 
  equatorial plane. It is of course important  to consider also their behavior over the full 
  $(R,z)$ plane, as 2D spectroscopy is nowadays routinely performed (e.g. Emsellem et al. 2007;
  Krajnovi\'c et al. 2008; Jeong et al. 2009).
  In Fig. \ref{fig:fig4} we show the two-dimensional maps of $\sigma$, $\sigphi$, and 
  $\vphib$ (for $k=1$), for a P11 model with $\eta=0.1$ and $\mu=0$; contours are displayed 
  for the full and the $\eta$-linear solutions. The comparison shows that the $\eta$-linear 
  $\sigma$ keeps extremely close to that of the full solution, even outside the equatorial plane; 
  a similar agreement persists for $\sigphi$, while it becomes slightly worse for $\vphib$. 
  However, even if  the shape of the isorotational surfaces in the $\eta$-linear approximation 
  seems more discrepant from that of the true solution than for the $\sigma$ and $\sigphi$ cases, 
  the $\vphib$ values of the $\eta$-linear and full solutions along cuts at fixed $z$ are still 
  very similar, as we verified with plots of these cuts (where indeed the differences in velocity 
  are of the same extent as  in the left panels of Fig. \ref{fig:fig2}).
  
  %%%%%%%%%%%%%%%%%%%%%%%%%%%%%%%%%%%%%%%%%%%%%%%%%%%%%%%%%%%%%%%%%%%%%%%
  \section{An application: rotation and flattening of globular clusters}\label{sec:NGC4372}
  %%%%%%%%%%%%%%%%%%%%%%%%%%%%%%%%%%%%%%%%%%%%%%%%%%%%%%%%%%%%%%%%%%%%%%%

  Globular Clusters (GCs) have traditionally been regarded as simple
  spherical, non-rotating stellar systems; however, small
  ellipticities have been observed since a long ago, and rotation is
  being detected in a growing number of them (e.g., Bianchini et
  al. 2018, Kamann et al. 2018, Ferraro et al. 2018). The origin of
  the observed flattening has been attributed to the effects of
  internal rotation, velocity dispersion anisotropy, and external
  tides (for a more extended discussion, see e.g. van
    den Bergh 2008). In particular, dynamical phenomena such as
    violent relaxation and two-body relaxation tend to produce
    isotropic velocity distributions in the central regions of stellar
    systems, so that, if flattening is observed there, rotation should
    be considered a possible explanation.  In addition to
  contributing to the shape of these systems, rotation is also
  expected to change their dynamical evolution (e.g., Fiestas et
  al. 2006), and to be linked to their `dynamical age' (e.g., Tiongco
  et al. 2017; Livernois et al.  2022, Leanza et al. 2022). Finally,
  rotation has been suggested to have a role in the formation of
  multiple stellar populations in them (Lacchin et
  al. 2023). Therefore, an assessment of the respective amounts of
  rotation and anisotropic pressure is particularly important. Indeed,
  in recent years much effort has been devoted to dynamical modelling
  of GCs, using different strategies, as for example $N$-body
  simulations (e.g., Hurley \& Shara 2012), Monte Carlo models (e.g.,
  Giersz et al. 2013; Kamlah et al. 2022), or self-consistent models
  specific for quasi-relaxed, rotating stellar systems (Varri \&
  Bertin 2012, Bianchini et al. 2013, Jeffreson et al. 2017); see
  Spurzem \& Kamlah (2023) for a recent review.
  
  In general, these techniques are quite complex, and their application time-consuming: it 
  would be desirable to have a simple but robust method to assess phenomenologically the 
  importance of rotation, before applying more sophisticated tools, and we suggest that the 
  homoeoidal expansion and the $\eta$-linear solutions of the Jeans equations could be one of such
  possibilities. Moreover, for the choice of Satoh's decomposition and for a density profile 
  roughly constant in the central regions, the homoeoidal expansion predicts a sort of 
  `universal profile' for the streaming velocity $\vphib$, of shape given by the first of equation
  \eqref{eq:Satoh} with $k=1$, coupled to equations \eqref{eq:rhos_Deltag} and 
  \eqref{eq:rhos_DeltaBH}. In particular, three main properties are predicted: 
  (1) from equation \eqref{eq:rhos_Deltag}, $\vphib$ scales as the square 
  root of flattening, and increases linearly with radius; 
  (2) it reaches a maximum;  
  (3) it decreases afterward.
  Of course these properties transfer also to the projected streaming velocity field $\vlos$. 
  Thus, a simple and direct relation between the shape of the system and its rotation profile 
  is expected, and it is tempting here to test whether it is satisfied by well observed systems. 
  At first sight, 
  the three features of $\vphib$ (and $\vlos$) agree with what observed, for a chosen test-case 
  object (see 
  below), and also for others (e.g., Leanza et al. 2022). Therefore, the method could provide 
  a fast and flexible tool to address, in a preliminary way, the following questions: are 
  observations consistent with velocity dispersion isotropy? if not, does a rescaling of 
  $\vlos$
  with a different costant $k$ value make the model consistent with observations? 
  or, is there the need for a change of $k$ with radius?
  
\begin{figure*}
        %\centering
        \includegraphics[width=0.5\linewidth]{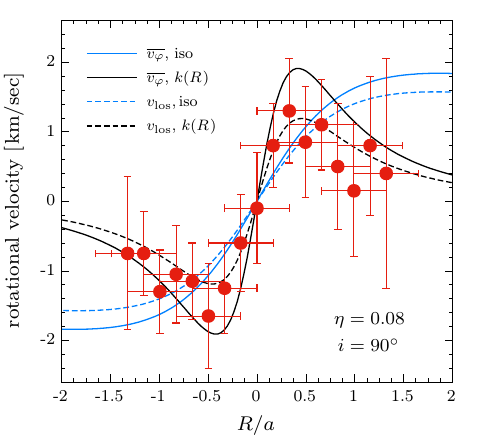}
        \hspace{-2mm}
        \includegraphics[width=0.5\linewidth]{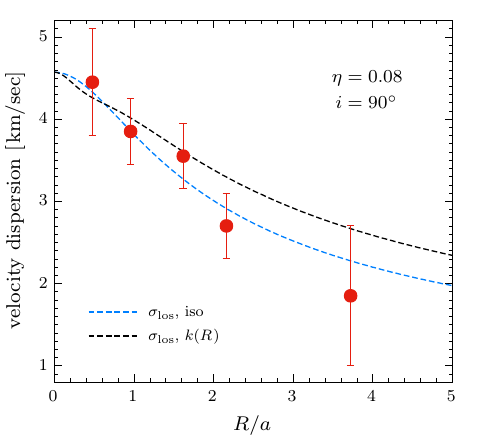}\\
        \vspace{0.5mm}
        \includegraphics[width=0.5\linewidth]{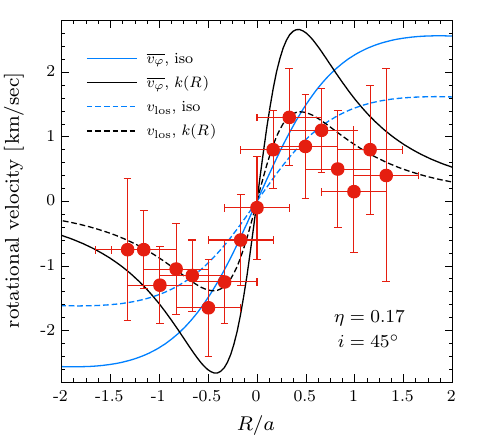}
        \hspace{-2mm}
        \includegraphics[width=0.5\linewidth]{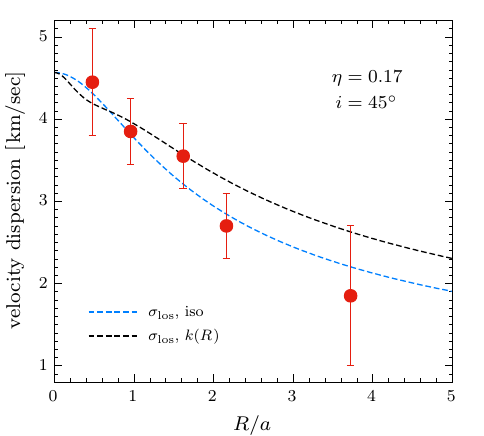}
        \vspace{2mm}
        \caption{The $\eta$-linear modelling of the Globular Cluster NGC 4372, whose observed 
          (projected along the line-of-sight) kinematics is shown by red points (from Kacharov et al. 2014).
          Solid lines show the intrinsic velocity $\vphib$; dashed lines show the projected velocity $\vlos$ (on the left) and 
          $\siglos$ (on the right). Blue lines show the isotropic case ($k=1$), black 
          lines the spatially-dependent $k(R\hspace{0.2mm})$ in equation \eqref{eq:k(R)}.
          Two inclination angles were adopted: $i=90^\circ$ (upper panels) and $i=45^\circ$ (lower panels). When 
                 $i=90^\circ$, the intrinsic flattening coincides with the observed one
                 ($\eta=0.08$); when $i=45^\circ$, the intrinsic flattening is $\eta=0.17$ 
                 (see Section \ref{sec:NGC4372}). }
  \label{fig:fig5}
  \end{figure*}

  As a test-case for the application of the homoeoidal method we chose NGC 4372, a GC for 
  which a detailed photometric and spectroscopic study was conducted (Kacharov et al. 2014). 
  NGC 4372 has an observed low ellipticity of $\eta=0.08$; and, thanks to a large number of
  precise radial velocity measurements, it has a $\vlos$ profile extending at least out to its 
  half-light radius\footnote{For a Plummer model, the characteristic radius $a$ corresponds 
  to the half-mass radius.}, and a velocity dispersion profile extending even further out. 
  Kacharov et al. (2014) adopted a Plummer model, one of the two illustrating cases above, 
  as an optimal representation of the observed properties; they estimated $a=5.1\,{\rm pc}$, 
  and $M=1.7\times\hspace{-0.2mm} 10^5\,M_\odot$. All this makes NGC 4372 an obvious candidate 
  for our test. We modelled then NGC 4372 with the P11 profile, of parameters as in Kacharov et 
  al., and,  based on the results of Section \ref{sec:linVSquad}, with the $\eta$-linear 
  solution of the Jeans equations. For the model, and for $k=1$, Figure \ref{fig:fig5} shows 
  the intrinsic streaming velocity $\vphib$ (blue solid curves), the line-of-sight  
  velocity $\vlos$ (blue dashed curves), and the line-of-sight velocity dispersion 
  $\siglos$ (see Section 5 in CMPZ21 and Chapter 11 in C21 for the formulae used 
  to obtain the projected quantities); the corresponding observed data points (red dots)
  are also shown for comparison, together with the their error bars. 
  When projecting, we adopted two inclination angles: 
  $i=90^\circ$ (upper panels in Fig. \ref{fig:fig5}) and $i=45^\circ$ (lower panels). In the 
  first case, NGC 4372 is supposed to be viewed edge-on, and the model was built with an intrinsic
  flattening coincident with the observed one ($\eta=0.08$); in the second case, the intrinsic
  flattening increases\footnote{
  When the line-of-sight is inclined by an angle $i$ with respect to the
  $z$-axis, the relation between the intrinsic 
  flattening $q$ and the observed flattening 
  $q_{\rm obs}$ is $q_{\rm obs}^2={\rm cos}^2 i+q^2{\rm sin}^2 i$ (e.g., see C21).} 
  to $\eta=0.17$. Overall, for both inclinations, $\siglos$ of the model accounts quite well for 
  the observed profile, but the isotropic $\vlos$ does not so: its innermost rising part does not 
  reproduce well the observed curve, and, more important, at distances larger than 
  $\simeq a$ it remains too high. We are then forced to exclude the possibility that NGC 4372 is 
  an isotropic rotator, and also that it is a rotator with a different but constant $k$, 
  that would have a $\vlos$ profile with the same shape, just rescaled. Notice that decreasing further 
  the inclination angle would not change significantly this conclusion: it would produce an 
  increased intrinsic
  flattening, and then an increase of the isotropic $\vphib$, that would be almost 
  perfectly compensated by the decrease of the projection angle.

  %%%%%%%%%%%%%%%%%%%%%%%%%%%%%%%%%%%%%%%%%%%%%%%%%%%%%%%%%%%%%%%%%%%%%%%%%%%%%%
  %%%%%%%%%%%%%%%%%%%%%%%%%%%%%%%%%%%%%%%%%%%%%%%%%%%%%%%%%%%%%%%%%%%%%%%%%%%%%%
  
  Having discarded the possibility of an isotropic rotator,
  we attempted then to reproduce the observed profile with a radially dependent Satoh 
  decomposition. Since the blue solid curves in Fig. \ref{fig:fig5}
  give the $\vphib$ field with $k=1$ in equation \eqref{eq:Satoh}, in practice they also 
  show $\sqrt{\Delta}$ and its projection; the modifications needed on $k$ can be then easily 
  deduced from these curves. Quite obviously,  we do not attach a deep physical meaning to 
  these modifications, even though some implications can be derived. An inspection of 
  Fig. \ref{fig:fig5}
  suggests that the required changes to $\vlos$, to be produced by a radially 
  dependent $k$, are:  
  (i) preserve the linear rise of $\vlos$ in the central regions, but include a sharp peak at 
  a radius of $\sim a/3$, that is not present in the constant $k$ case; 
  (ii) be significantly lower than the isotropic rotation velocity outside $\simeq a$.
  We parametrized these requests with the trial function
  \begin{equation}
      k(R\hspace{0.2mm})=
      \frac{A}{1+(\Rtil/B)^n},
      \label{eq:k(R)}
  \end{equation}
  where $\Rtil=R/a$, and $A$, $B$ and $n$ are three dimensionless free
  parameters.  In Fig. \ref{fig:fig5} with black lines we show the
  intrinsic and projected streaming velocity profiles, obtained from
  equation \eqref{eq:k(R)}, with $A=3.5$, $B=0.5$, and $n=2$, and for
  the two inclination angles $i=90^\circ$ and $i=45^\circ$. The chosen
  values of $A$, $B$ and $n$ are not the result of a rigorous ``best
  fitting'' procedure; they reproduce quite reasonably the observed
  velocity profile, and allow us to draw three robust conclusions: the
  central regions must rotate faster than the isotropic rotator, as
  $k \simeq 3.5$ there; rotation is very concentrated; and the
  $\vphib$ decline for $R \gtrsim a/2$ is steep, with
  $k \propto 1/R^2$.  The lack of proper motion
    measurements for NGC 4372 prevented us from establishing the
    inclination angle, thus the intrinsic flattening. It would be
    interesting to extend our analysis to some other GCs
    with well-measured proper motions; however, as stressed above, we
    found a compensation between the system inclination and $\vlos$,
    therefore we are confident that the results obtained are quite
    robust.

  The right panels of Fig. \ref{fig:fig5} also show that, with the $k(R)$ in equation 
  \eqref{eq:k(R)}, $\siglos$ 
  differs from that of the isotropic rotator, which is not a surprise because $\vlos$ enters the
  expression for $\siglos$ (see e.g. equation 54 in CMPZ21): this is at the origin
  of the (small) drop of the black lines in the very central regions. In particular, the two outermost
  data points are better reproduced by the isotropic $\siglos$ rather than the 
  new one. We believe that a formal solution, reproducing both $\vlos$ and $\siglos$, 
  could be obtained by using a more complicated functional form of
  $k(R)$, for example that increases again up to unity outer of the most external observed 
  point of $\vlos$; 
  however, we consider this possibility quite implausible from a physical point of view. 
  We conclude that NGC 4372 is unlikely to 
  be an isotropic rotator, because of its lower rotation at $R\gtrsim a$, and a higher rotation in 
  its central region. Reassuringly, some of these conclusions have been also reached with a 
  more sophisticated approach, based on the construction of models supported by a self-consistent 
  phase-space distribution function (e.g. Varri \& Bertin 2012, Jeffreson et al. 2017).
  
  Before concluding this analysis, it is tempting to suggest another possible interpretation for the
  observed kinematic features of NGC 4372: the GC could be a two-component system, with an 
  inner rotating 
  structure physically distinct from that of the main body of the GC, and described by its own 
  phase-space distribution function. Our modelling so far was implicitly based on the use of a 
  single distribution function, i.e., the GC was assumed to be a one-component system. If the 
  total  distribution function were the sum of two different distribution functions, one for the 
  non-rotating (or slowly rotating) GC, and the other for the fast rotating substructure, the total
  rotational field to be modelled with the Jeans equations were the mass averaged rotational field 
  of the GC and of the substructure (not just that of the sampled stars of the subcomponent). It 
  would be interesting to determine observationally if the stars contributing to the projected 
  streaming velocity in the central region show a difference in age and/or chemical composition 
  with respect to the majority of the stars of the GC.
  
  A different possibility would be that the rotational
    profile is explained by a significant change in the flattening of
    the system approaching the centre; in fact the ellipticity is
    observed to vary in the central regions of some GC (e.g.
    Bianchini et al. 2013). The possibility that the inner regions can
    be actually interpreted as a flattened isotropic rotator is
    qualitatively supported by the scaling of the isotropic $\vphib$
    with $\sqrt{\eta}$. We note however that in NGC 4372 the
    fiducial value $\simeq 3$ of the Satoh $k$ parameter in the
    central regions would require, if decreased to 1, an increase of
    the adopted $\eta$ by a factor of $\approx 9$, bringing the
    flattening well above the limiting value allowed by the homoeoidal
    expansion.
  
%%%%%%%%%%%%%%%%%%%%%%%%%%%%%%%%%%%%%%%%%%%%%%%%%%%%%%%%%%%%%%%%%%%%%%
  \section{Discussion and conclusions}\label{sec:Conclusions}
%%%%%%%%%%%%%%%%%%%%%%%%%%%%%%%%%%%%%%%%%%%%%%%%%%%%%%%%%%%%%%%%%%%%%%

  In this work we studied some aspects of the Jeans modelling of axisymmetric systems 
  that are only slightly deviating from a spherical shape, a situation often encountered in 
  applications. In particular, we considered two problems related to the homoeoidal expansion
  technique (CB05; CMPZ21; C21; see also Lee and Suto 2003; Muccione and Ciotti 2004; Ciotti et 
  al. 2006; Ciotti and Pellegrini 2008). This technique allows for a simple modelling of systems
  sligthly departing from spherical symmetry, based on the 
  expansion of the original ellipsoidal density-potential pair at the linear order in terms of 
  the density flattening $\eta$. Thanks to this expansion, a numerical integration for the 
  determination 
  of the potential can be usually avoided, and the resulting (two-integral) Jeans equations can 
  often be solved analytically. Even in case of a numerical treatment, the integrals are no more 
  difficult than for spherically symmetric models.
  
  Two interesting questions concerning the homoeoidal expansion, especially relevant 
  in modelling
  applications, were not properly addressed so far. The first is related to the physical 
  interpretation of the expanded density-potential pair, which obeys exactly the Poisson equation, 
  and that can be interpreted as the linearization of the original ellipsoidal models, or as a 
  genuinely self-consistent model. In the first interpretation, 
  only linear terms in the flattening $\eta$ are retained in the solutions of the Jeans equations 
  ($\eta$-linear solutions), while in the second interpretation all terms up to the quadratic 
  order are considered ($\eta$-quadratic solutions). The question is then to estimate the 
  contribution of these quadratic terms to the solutions (even in light of the fact that 
  such terms do not present special mathematical difficulties in the analytical treatment).  
  The problem is not of secondary importance as it might appear: even if $\eta^2$ is much smaller 
  than $\eta$ for small values of $\eta$, it is not guaranteed that the corresponding 
  coordinate-dependent functional coefficients in the expansion are necessarily small, and so  
  the discarded $\eta$-quadratic terms could be non negligible over some region of space. 
  The second question is related to the additional fact that the 
  $\eta$-quadratic solutions of the Jeans equations are {\it not} the quadratic truncation of 
  the expansion of the full solutions in terms of powers of the flattening. 
  Therefore, it is interesting to estimate not 
  only how the $\eta$-linear and $\eta$-quadratic solutions differ, but also how they deviate 
  from the full solution. To quantitatively answer the questions above, we obtained the 
  analytical $\eta$-quadratic solutions, and the (numerical) solution 
  of the two-integral Jeans equations, for two weakly flattened ellipsoidal systems, namely 
  the ellipsoidal Plummer model and the Perfect Ellipsoid. We found that, for flattening of 
  the order of $\eta = 10^{-1}$, the differences between the $\eta$-linear and $\eta$-quadratic 
  solutions are everywhere negligible; moreover, the $\eta$-linear solution 
  already provides an excellent agreement with the full solution, and then suffices for practical
  purposes.
  
  For an example of application of the use of the $\eta$-linear solution, we chose the
  research field of GCs, systems with small flattening often described by the Plummer model.  
  The comparison with GCs was also suggested by the fact that the isotropic streaming velocity 
  field of weakly flattened ellipsoidal systems is in general linearly rising in the inner 
  part (it scales as the square root of the flattening, i.e. $\sqrt{\eta}$), it reaches a maximum, and then shows a 
  monotonic decline; this behaviour is remarkably similar to the phenomenological velocity 
  profile usually 
  adopted to describe the rotation of GCs. We considered then the GC NGC 4372, characterized 
  by a small flattening ($\eta=0.08$), and with an extended rotation curve observed. 
  Our modelling rules out the possibility that NGC 4372 is an isotropic stellar system flattened 
  by rotation, in agreement with the conclusions obtained by using more sophisticated modelling
  techniques, for example based on the construction of self-consistent solutions starting from 
  the phase-space distribution function (e.g. Varri \& Bertin 2012; Bianchini et al. 2013;
  Jeffreson et al. 2017). Interestingly, we show that rotation must exceed that of an isotropic 
  rotator in the central region, which indicates the possibility of the presence a separate 
  highly rotating
  subcomponent. We conclude that the $\eta$-linear homoeoidally expanded solutions can be a useful
  starting point to gain insight into the internal dynamics of weakly flattened and rotating 
  stellar systems (as some GCs) before turning to more complex studies.

%%%%%%%%%%%%%%%%%%%%%%%%%%%%%%%%%%%%%%%%%%%%%%%%%%%%%%%%%%%%%%%%%%%%%%%%%%%%%%%%
  \section*{Acknowledgements}
%%%%%%%%%%%%%%%%%%%%%%%%%%%%%%%%%%%%%%%%%%%%%%%%%%%%%%%%%%%%%%%%%%%%%%%%%%%%%%%%

  We thank the anonymous Referee for important comments and useful
  suggestions that improved the paper content and presentation.

%%%%%%%%%%%%%%%%%%%%%%%%%%%%%%%%%%%%%%%%%%%%%%%%%%%%%%%%%%%%%%%%%%%%%%%%%%%%%%%%
  \section*{Data Availability}
%%%%%%%%%%%%%%%%%%%%%%%%%%%%%%%%%%%%%%%%%%%%%%%%%%%%%%%%%%%%%%%%%%%%%%%%%%%%%%%%

  No datasets were generated or analysed in support of this research.

%%%%%%%%%%%%%%%%%%%%%%%%%%%%%%%%%%

%%%%%%%%%%%%%%%%%%%%%%%%%%%%%%%%%%%%%%%%%%%%%%%%%%%%%%%%%%%%%%%%%%%%%%%%%%%%%%%%%%%%%%%%%%%%%%
%***************************************APPENDICI*********************************************
%%%%%%%%%%%%%%%%%%%%%%%%%%%%%%%%%%%%%%%%%%%%%%%%%%%%%%%%%%%%%%%%%%%%%%%%%%%%%%%%%%%%%%%%%%%%%%

  \appendix

%\begin{comment}
  \onecolumn
  \section{Plummer Model}\label{app:Plummer}
  For the ellipsoidal generalization of the Plummer model, the three dimensionless functions
  in equation \eqref{eq:rho_Psi} are 
  \begin{equation}
        \rho_0(s)=\frac{3}{(1+s^2)^{5/2}}\hspace{0.3mm},\hspace{6mm}
        \rho_1(s)=\frac{3\hspace{0.2mm}(1-4s^2)}{(1+s^2)^{7/2}}\hspace{0.3mm},\hspace{6mm}
        \rho_2(s)=\frac{15}{(1+s^2)^{7/2}}\hspace{0.3mm}.
  \label{eq:rhoti_P11}
  \end{equation}
  The associated dimensionless potentials in equation \eqref{eq:rho_Psi} can be easily obtained:
  \begin{equation}
        \Psi_0(s)=\frac{1}{(1+s^2)^{1/2}}\hspace{0.3mm},\hspace{6mm}
        \Psi_1(s)=\frac{3s^2+2}{s^2(1+s^2)^{3/2}}-\frac{2\hspace{0.3mm}\arcsh s}{s^3}\hspace{0.3mm},\hspace{6mm}
        \Psi_2(s)=-\hspace{0.4mm}\frac{4s^2+3}{s^4(1+s^2)^{3/2}}+\frac{3\hspace{0.3mm}\arcsh s}{s^5}\hspace{0.3mm},
  \end{equation}
  so that the two components of the circular velocity in the
  $\eta$-linear expansion \eqref{eq:vcirc_exp} are
  \begin{equation}
        \varv_0^2(\Rtil\hspace{0.15mm})=\frac{\Rtil^2}{(1+\Rtil^2)^{3/2}}\hspace{0.3mm},\hspace{6mm}
        \varv_1^2(\Rtil\hspace{0.15mm})=\frac{3\hspace{0.3mm}\arcsh \Rtil}{\Rtil^3}-\frac{4\Rtil^2+3}{\Rtil^2(1+\Rtil^2)^{3/2}}\hspace{0.3mm}.
  \label{eq:vc_P11}
  \end{equation}
  
  The three functions in equation \eqref{eq:Xi} describing the BH contribution to the vertical velocity 
  dispersion of the stars are
\begin{equation}
Y_0(s)=
\frac{8s^4+12s^2+3}{s\hspace{0.2mm}(1+s^2)^{3/2}}-8\hspace{0.3mm},
\qquad\,\,
Y_1(s)=
\frac{16s^6+40s^4+30s^2+3}{s\hspace{0.2mm}(1+s^2)^{5/2}}-16\hspace{0.3mm},
\qquad\,\,
Y_2(s)=
\frac{3\hspace{0.1mm}(16s^6+40s^4+30s^2+5)}{s\hspace{0.2mm}(1+s^2)^{5/2}}-48\hspace{0.2mm}.
\end{equation}

For the contribution of the galaxy model to the velocity dispersion in equation \eqref{eq:Hij},
an elementary integration shows that
\begin{equation}
\Hzz(s)=\frac{1}{(1+s^2)^3}\hspace{0.2mm},
\end{equation}
\begin{equation}
\Hzu(s)=
\frac{3\hspace{0.2mm}(8s^8+27s^6+31s^4+15s^2+2)}{s^2(1+s^2)^4}
+\frac{6\hspace{0.2mm}(2s^2+1)(8s^4+8s^2-1)}{s^3(1+s^2)^{3/2}}
\hspace{0.4mm}\arcsh s-48\hspace{0.2mm}L(s)\hspace{0.3mm},
\end{equation}
\begin{equation}
\Hzd(s)=
\frac{96s^{10}+324s^8+380s^6+167s^4+3s^2-9}{s^4(1+s^2)^4}
+\frac{3\hspace{0.2mm}(128s^8+192s^6+48s^4-8s^2+3)}{s^5(1+s^2)^{3/2}}
\hspace{0.4mm}\arcsh s-192\hspace{0.2mm}L(s)\hspace{0.3mm},
\end{equation}
\begin{equation}
\Huu(s)
=\frac{384s^{10}+1680s^8+2816s^6+2119s^4+641s^2+24}{2s^2(1+s^2)^5}
+\frac{12\hspace{0.2mm}(64s^8+160s^6+120s^4+20s^2-1)}{s^3(1+s^2)^{5/2}}
\hspace{0.4mm}\arcsh s-384\hspace{0.2mm}L(s)\hspace{0.3mm},
\end{equation}
\begin{equation}
\begin{aligned}
\Hud(s)&=
\frac{768s^{12}+3360s^{10}+5632s^8+4361s^6+1438s^4+99s^2-9}{s^4(1+s^2)^5}\\[5pt]
&+\frac{3\hspace{0.2mm}(1024s^{10}+2560s^8+1920s^6+320s^4-40s^2+3)}{s^5(1+s^2)^{5/2}}
\hspace{0.4mm}\arcsh s-1536\hspace{0.2mm}L(s)\hspace{0.3mm},
\end{aligned}
\end{equation}
\begin{equation}
\begin{aligned}
\Hdd(s)&=
\frac{3840s^{12}+16800s^{10}+28160s^8+21805s^6+6983s^4+180s^2-180}{2s^4(1+s^2)^5}\\[5pt]
&+\frac{30\hspace{0.2mm}(2s^2+1)(128s^8+256s^6+112s^4-16s^2+3)}{s^5(1+s^2)^{5/2}}
\hspace{0.4mm}\arcsh s-3840\hspace{0.2mm}L(s)\hspace{0.3mm},
\end{aligned}
\end{equation} 
where $L(s) \equiv \ln 4+\ln(1+s^2)$.
As in the text, $\Rtil=R/a$, and $s=r/a$, where $a$ is the scale length of the model.
  
  \section{Perfect Ellipsoid}\label{app:Perf_Ellips}
  
  For the Perfect Ellipsoid model, the three dimensionless functions in equation \eqref{eq:rho_Psi} are
  \begin{equation}
        \rho_0(s)=\frac{4}{\upi\hspace{0.2mm}(1+s^2)^2}\hspace{0.3mm},\hspace{6mm}
        \rho_1(s)=\frac{4\hspace{0.2mm}(1-3s^2)}{\upi\hspace{0.2mm}(1+s^2)^3}\hspace{0.3mm},\hspace{6mm}
        \rho_2(s)=\frac{16}{\upi\hspace{0.2mm}(1+s^2)^3}\hspace{0.1mm}.
  \label{eq:rhoti_ZLM85}
  \end{equation} 
  The associated dimensionless potentials in equation \eqref{eq:rho_Psi} can be easily obtained:
  \begin{equation}
        \Psi_0(s)=\frac{2\arctan s}{\upi s}\hspace{0.3mm},\hspace{6mm}
        \Psi_1(s)=\frac{4\arctan s}{\upi s^3}-\frac{2\hspace{0.1mm}(s^2+2)}{\upi s^2(1+s^2)}\hspace{0.3mm},\hspace{6mm}
        \Psi_2(s)=-\hspace{0.4mm}\frac{6\arctan s}{\upi s^5}+\frac{2\hspace{0.1mm}(2s^2+3)}{\upi s^4(1+s^2)}\hspace{0.3mm},
  \end{equation}
  so that the two components of the circular velocity in the
  $\eta$-linear expansion are:
  \begin{equation}
        \varv_0^2(\Rtil\hspace{0.15mm})=\frac{2\arctan \Rtil}{\upi \Rtil}-\frac{2}{\upi(1+\Rtil^2)}\hspace{0.3mm},\hspace{6mm}
        \varv_1^2(\Rtil\hspace{0.15mm})=\frac{2\hspace{0.1mm}(2\Rtil^2+3)}{\upi\Rtil^2(1+\Rtil^2)}-\frac{6\arctan \Rtil}{\upi \Rtil^3}\hspace{0.3mm}.
  \label{eq:vc_ZL85}
  \end{equation}
  
  The three functions in equation \eqref{eq:Xi} describing the BH contribution to the vertical velocity 
  dispersion of the stars are
\begin{equation}
Y_0(s)=
\frac{6\arctan s}{\upi}+\frac{2\hspace{0.1mm}(3s^2+2)}{\upi s\hspace{0.2mm}(1+s^2)}-3\hspace{0.3mm},
\qquad\,
Y_1(s)=
\frac{12\arctan s}{\upi}+\frac{4\hspace{0.1mm}(3s^4+5s^2+1)}{\upi s\hspace{0.2mm}(1+s^2)^2}-6\hspace{0.3mm},
\qquad\,
Y_2(s)=
-\hspace{0.4mm}\frac{30\hspace{0.4mm}\arccot\,s}{\upi}+\frac{2\hspace{0.1mm}(15s^4+25s^2+8)}{\upi s\hspace{0.2mm}(1+s^2)^2}.
\end{equation}

For the contribution of the galaxy model to the velocity dispersion in equation \eqref{eq:Hij},
an elementary integration shows that
\begin{equation}
\Hzz(s)=
\frac{12\arctan^2\hspace{-0.3mm} s}{\upi^2}
+\frac{8\hspace{0.2mm}(3s^2+2)}{\upi^2 s\hspace{0.2mm}(1+s^2)}\arctan s
+\frac{4\hspace{0.2mm}(3s^2+4)}{\upi^2(1+s^2)^2}
-3\hspace{0.3mm},
\end{equation}
\begin{equation}
\Hzu(s)=
-\hspace{0.4mm}\frac{48\arctan^2\hspace{-0.3mm} s}{\upi^2}
-\frac{8\hspace{0.2mm}(12s^6+20s^4+7s^2-2)}{\upi^2 s^3(1+s^2)^2}\arctan s
-\frac{8\hspace{0.2mm}(6s^6+14s^4+9s^2+2)}{\upi^2 s^2(1+s^2)^3}
+12\hspace{0.3mm},
\end{equation}
\begin{equation}
\Hzd(s)=
-\hspace{0.4mm}\frac{180\arctan^2\hspace{-0.3mm} s}{\upi^2}
-\frac{8\hspace{0.2mm}(45s^8+75s^6+24s^4-4s^2+3)}{\upi^2 s^5(1+s^2)^2}\arctan s
-\frac{4\hspace{0.2mm}(45s^8+105s^6+68s^4+4s^2-6)}{\upi^2 s^4(1+s^2)^3}
+45\hspace{0.3mm},
\end{equation}
\begin{equation}
\Huu(s)=
-\hspace{0.4mm}\frac{480\arctan^2\hspace{-0.3mm} s}{\upi^2}
-\frac{32\hspace{0.2mm}(30s^6+50s^4+16s^2-1)}{\upi^2 s^3(1+s^2)^2}\arctan s
-\frac{16\hspace{0.2mm}(90s^8+300s^6+343s^4+136s^2+6)}{3\hspace{0.2mm}\upi^2 s^2(1+s^2)^4}
+120\hspace{0.3mm},
\end{equation}
\begin{equation}
\Hud(s)=
-\hspace{0.4mm}\frac{1680\arctan^2\hspace{-0.3mm} s}{\upi^2}
-\frac{8\hspace{0.2mm}(420s^8+700s^6+224s^4-32s^2+3)}{\upi^2 s^5(1+s^2)^2}\arctan s
-\frac{8\hspace{0.2mm}(630s^{10}+2100s^8+2422s^6+1036s^4+81s^2-9)}{3\hspace{0.2mm}\upi^2 s^4(1+s^2)^4}
+420\hspace{0.3mm},
\hspace{0.4mm}
\end{equation}
\begin{equation}
\Hdd(s)=
-\hspace{0.4mm}\frac{3780\arctan^2\hspace{-0.3mm} s}{\upi^2}
-\frac{24\hspace{0.2mm}(315s^8+525s^6+168s^4-24s^2+8)}{\upi^2 s^5(1+s^2)^2}\arctan s
-\frac{4\hspace{0.2mm}(945s^{10}+3150s^8+3633s^6+1536s^4+64s^2-48)}{\upi^2 s^4(1+s^2)^4}
+945\hspace{0.3mm}.
\end{equation} 
As in the text, $\Rtil=R/a$, and $s=r/a$, where $a$ is the scale length of the model.
%%%%%%%%%%%%%%%%%%%%%%%%%%%%%%%%%%%%%%%%%%%%%%%%%%%%%%%%%%%%%%%%%%%%%%%%%%%%%%%%%%%%%%%%%%%%%

\end{document}